\documentstyle[12pt,epsfig]{article} 
\newcommand{\pr}[1]{Phys. Rev. #1}

\def\beq{\begin{equation}}
\def\eeq{\end{equation}}
\def\bea{\begin{eqnarray}}
\def\eea{\end{eqnarray}}
\def\bq{\begin{quote}}
\def\eq{\end{quote}}
\def\ben{\begin{enumerate}}
\def\een{\end{enumerate}}

\def\ie{{\it i.e.}}

\def\lesssim{\mathrel{\mathpalette\vereq<}}
\def\gtrsim{\mathrel{\mathpalette\vereq>}}
\makeatletter
\def\vereq#1#2{
\lower3pt\vbox{\baselineskip1.5pt \lineskip1.5pt
\ialign{$\m@th#1\hfill##\hfil$\crcr#2\crcr\sim\crcr}}}
\makeatother

\def\beq{\begin{equation}}
\def\eeq{\end{equation}}
\def\bea{\begin{eqnarray}}
\def\eea{\end{eqnarray}}
\def\bq{\begin{quote}}
\def\eq{\end{quote}}
\def\ben{\begin{enumerate}}
\def\een{\end{enumerate}}

\begin{document}

\begin{titlepage}
\begin{center}
\hfill    FERMILAB-Pub-02/066-T\\
\hfill nuhep-exp/2002-01\\
~{} \hfill hep-ph/0204208\\

\vskip 1cm

{\large \bf Neutrino Oscillations with a Proton Driver Upgrade and
an Off-Axis Detector: A Case Study}

\vskip 1cm

Gabriela Barenboim, Andr\' e de Gouv\^ ea,

\vskip 0.5cm

{\em Theoretical Physics Division, Fermilab \\ 
 P.O. Box 500, Batavia, IL 60510, USA}

\vskip 1cm

Micha{\l} Szleper, and Mayda Velasco

\vskip 0.5cm

{\em Northwestern University  \\                   
Department of Physics \& Astronomy \\              
2145 Sheridan Rd, Evanston, IL 60208, USA} 

\end{center}

\vskip 2cm

\begin{abstract}

We study the physics capabilities of the NuMI beamline with an off-axis highly-segmented
iron scintillator detector and with the inclusion of the currently under study 
proton driver upgrade.
We focus on the prospects for the experimental determination of the
remaining neutrino oscillation parameters, assuming different outcomes for experiments
under way or in preparation.  An optimization
of the beam conditions and detector location for the detection of
the $\nu_{\mu} \rightarrow \nu_e$-transitions is discussed. 
Different physics scenarios were considered, depending on the actual solution of the
solar neutrino puzzle.  If KamLAND measures $\Delta m^2_{\odot}$, we
find it possible to measure both $|U_{e3}|^2$ and the CP violating
phase $\delta$ within a viable exposure time, assuming a realistic
detector and a complete data analysis.  Exposure to both neutrino
and antineutrino beams is necessary.  We can, in addition, shed
light on $\Delta m^2_{\odot}$ if its value is at the upper limit of
KamLAND sensitivity ({\it i.e.,}\/ the precise value of $\Delta m^2_{\odot}$ remains
unknown even after KamLAND).  If the solar
neutrino solution is not in the  LMA region, we can measure $|U_{e3}|^2$ and
determine the neutrino mass hierarchy.  The existence of the proton
driver is vital for the feasibility of most of these measurements.   

\end{abstract}

\vskip 2cm
\end{titlepage}

\section{Introduction}

There is hope that the reconstruction
of the neutrino mass matrix will shed light on some of the most relevant
open questions faced by high energy physics today, including 
the origin of the neutrino mass, the flavor-puzzle,
and the origin of the asymmetry between matter and antimatter, among others.

In order to start tackling these issues,  ``precision measurements''
of the neutrino parameters are required: simply knowing 
that the neutrino masses are tiny is not enough. Knowing that there are
three non degenerate massive neutrinos is also not enough to prove that there
is CP violation in the lepton sector, although it would be rather 
surprising if it turned out otherwise. 
Forthcoming neutrino experiments, therefore, have to be designed
not simply to test the oscillation hypothesis, 
but also to perform precise measurements of the oscillation parameters.

%We start by describing where we are now and what
%would be interesting to measure from the generation of experiments to follow 
%K2K, Kamland,  MINOS and CNGS. 

We currently know that there are (at least) 
three neutrino ``flavors'' -- the electron, the muon and the tau neutrinos --
and the current neutrino oscillation data strongly suggest 
that the flavor eigenstates differ from the mass eigenstates  and 
therefore the neutrinos can mix. Furthermore, the atmospheric, solar and
reactor data point to two hierarchically different mass-squared differences
plus a small ``connecting angle,'' such that, to good approximation, both the
atmospheric and solar neutrino puzzles can be solved by assuming
$\nu_{\mu}\leftrightarrow\nu_{\tau}$-oscillations and $\nu_e\leftrightarrow\nu_{\rm
other}$-oscillations, respectively. 
Forthcoming results from  solar, atmospheric, accelerator, and
reactor neutrino oscillation experiments will measure 
the size of the mass-squared difference and mixing angle that drive the
solar neutrino oscillations with good precision and 
%(and depending on the solution 
%we would be able to pin down its value), 
the mass-squared difference and the
mixing angle involved in the atmospheric oscillation to about
10\% \cite{bar}
(and there should be direct confirmation that muon neutrinos oscillate
to tau neutrinos by direct observation of a tau appearance
\cite{CNGS}). Furthermore, we will also definitively 
open (or close) the door to extra new 
physics in the neutrino sector by confirming (or not)
the LSND anomaly \cite{LSND} with MiniBooNE data \cite{miniboone}.\footnote{We 
will not consider the LSND anomaly in this paper.} 

In addition, some non-oscillation neutrino experiments are aiming at 
determining the absolute value of the neutrino masses using direct 
searches~\cite{katrin} 
and the nature of the neutrino mass: are they Majorana or Dirac 
particles?~\cite{doubleb}.  The knowledge of the neutrino mass 
is important  to  decide whether neutrinos contribute
significantly to the total energy of the Universe.

In view of the above results and the ongoing experimental programs,
the ultimate goal for the next generation of neutrino experiments should be
to test CP violation in the neutrino sector, if the parameters of the solar
neutrino solution allow such a determination. One way to achieve this goal is
to devise experiments that:
\begin{itemize}
        \item are sensitive to the sub-dominant $\nu_\mu \rightarrow \nu_e$
        channel for the atmospheric $L/E_\nu$,  and can measure $|U_{e3}|$ from
        both   {\mbox{$\nu_\mu \rightarrow \nu_e$}} and
        {\mbox{$\bar{\nu}_\mu \rightarrow \bar{\nu}_e$}} transitions;
        \item  are capable of determining the neutrino mass pattern
          from matter effects; 
	\item ultimately, can test CP invariance in the leptonic sector.
\end{itemize}

Such experiments require, in general, relatively intense and mono-energetic beams.
As first suggested in \cite{e889}, this could be achieved with off-axis 
neutrino beams.

The literature contains a significant amount of work 
\cite{bar2, bar3} which discusses the potential of generic 
``super-beams,'' with neutrino energies ranging from 1 to 50 GeV and baselines
spanning from 200 to 7000 kilometers. Low energy neutrino beams
have also been studied \cite{jj}.
In this work, we explore which of the questions above 
can be addressed by using the NuMI beam line with a new proton driver
and the construction of an off-axis detector. We will concentrate exclusively
on what one may hope to learn by looking for 
$\nu_{\mu}\rightarrow\nu_e$ appearance in a highly 
segmented iron-scintillator 
detector.  We will conclude, in agreement with some of these  
previous works, that intense beams can significantly improve our knowledge of
the neutrino oscillation parameters, including (depending on the solar solution)
some sensitivity to a CP violating phase. However, the ultimate 
sensitivity to some of the neutrino parameters, in particular the CP
violating phase, will require the purity and intensity of neutrino factory beams
\cite{fac}.

The paper is organized as follows.  First, we briefly review the 
neutrino mixing matrix and oscillation probabilities, and describe the
current knowledge of neutrino masses and mixing angles. In Sec.~3,
we describe the off-axis neutrino beam, and discuss where an off-axis detector 
should be located in order to maximize its physics capabilities. 
In Sec.~4, we describe in detail the detector we will be considering,
and discuss reconstruction efficiencies and strategies for reducing the number 
of background events.  In Sec.~5, we discuss the physics capabilities of such 
a setup, for different values of the solar mass-squared difference. We find 
that while some information can be obtained from a ``neutrino'' beam, it is 
imperative to run with an ``antineutrino'' beam as well. Furthermore, in order 
to extract more ``exciting'' physics (such as CP-violation) out of the 
experiment, a new intense proton source and a large detector 
are required. In Sec.~6 we summarize our results and compare our findings
with similar studies at different beamlines.

\section{Neutrino Mixing and Oscillations}
\label{theory_sec}

The presence of non-zero masses for the light neutrinos introduces a leptonic
mixing matrix, $U$, analogous to the well known CKM quark mixing matrix, 
and which in general is not expected to be diagonal. 
The matrix $U$ connects the neutrino flavor eigenstates with the mass eigenstates: 

\begin{equation}
	|\nu_\alpha\rangle = \sum_i U_{\alpha i}|\nu_i\rangle,
\end{equation}

\noindent
where $\alpha$ denotes the active neutrino flavors, $e,\ \mu$ or 
$\tau$, while $i$ runs over the mass eigenstates.  
%We will choose to parameterize $U$ in the standard PDG form \cite{PDG}, namely,
%%
%\begin{equation}
%U
%= \left( \begin{array}{ccc}
%  c_{13} c_{12}       & c_{13} s_{12}  & s_{13} e^{-i\delta} \\
%- c_{23} s_{12} - s_{13} s_{23} c_{12} e^{i\delta}
%& c_{23} c_{12} - s_{13} s_{23} s_{12} e^{i\delta}
%& c_{13} s_{23} \\
%    s_{23} s_{12} - s_{13} c_{23} c_{12} e^{i\delta}
%& - s_{23} c_{12} - s_{13} c_{23} s_{12} e^{i\delta}
%& c_{13} c_{23} \\
%\end{array} \right) \,,
%\end{equation}
%
%where $c_{jk} \equiv \cos\theta_{jk}$ and $s_{jk} \equiv \sin\theta_{jk}$.
It is ``traditional'' to define
the mixing angles $\theta_{12,13,23}$ in the following way:
\begin{equation}
\tan^2\theta_{12}\equiv \frac{|U_{e2}|^2}{|U_{e1}|^2},~~~
\tan^2\theta_{23}\equiv \frac{|U_{\mu3}|^2}{|U_{\tau3}|^2},~~~
\sin^2\theta_{13}\equiv |U_{e3}|^2,
\end{equation} 
while 
\begin{equation}
\Im(U_{e2}^*U_{e3}U_{\mu2}U_{\mu3}^*)\equiv 
\sin\theta_{12}\cos\theta_{12}\sin\theta_{23}\cos\theta_{23} 
\sin\theta_{13}\cos^2\theta_{13}\sin\delta, 
\end{equation}
defines the CP-odd phase $\delta$. 
For Majorana neutrinos, $U$ contains two further multiplicative phase
factors, but these are invisible to oscillation phenomena.

In order to relate the mixing angles and mass-squared differences
to the parameters constrained by experiments, it is convenient to define
the neutrino masses such that $m_1^2<m^2_2$  with 
$\Delta m^2_{12}<|\Delta m^2_{13,23}|$\footnote{We define 
$\Delta m^2_{ij}\equiv m^2_j -m^2_i$.} (the data, in fact, point to 
$\Delta m^2_{12}\ll|\Delta m^2_{13,23}|$). With this definition, the ``solar
angle'' $\theta_{\odot}\simeq\theta_{12}$, while the ``atmospheric angle''
$\theta_{\rm atm}\simeq\theta_{23}$. Furthermore, reactor experiments 
constrain $|U_{e3}|^2$. The solar mass-squared difference
$\Delta m^2_{\odot}=\Delta m^2_{12}$, while the atmospheric mass-squared
difference is $\Delta m^2_{\rm atm}=|\Delta m^2_{13}|\simeq|\Delta m^2_{23}|$.
It is important to note that $m_3^2$ can be either larger or smaller
than $m_1^2,m_2^2$.  

The oscillation probability $P(\nu_\alpha \rightarrow \nu_\beta)$ 
is given by the absolute square of the overlap of 
the observed flavor state, $|\nu_\beta\rangle$, with the time-evolved
initially-produced flavor state, $|\nu_\alpha\rangle$.  In vacuum,  
it yields the well-known result:
\begin{equation}
\begin{array}{rl}
	P(\nu_\alpha \rightarrow \nu_\beta) =&\left|\langle\nu_\beta | 
		e^{-iH_0L}|\nu_\alpha\rangle\right|^2 
	      =	\sum_{i,j} U_{\alpha i}U^*_{\beta i}U^*_{\alpha j}U_{\beta j}
		e^{-i\Delta m^2_{ij}L/2E}\\[0.1in]
	=&P_{\rm CP-even}(\nu_\alpha \rightarrow \nu_\beta) 
		+ P_{\rm CP-odd}(\nu_\alpha \rightarrow \nu_\beta) \; . \\[0.1in]
\end{array}
\end{equation}
\noindent
The CP-even and CP-odd contributions are
\begin{equation}
\begin{array}{rl}
	P_{\rm CP-even}(\nu_\alpha \rightarrow \nu_\beta) =&P_{\rm CP-even}(
		\bar{\nu}_\alpha \rightarrow \bar{\nu}_\beta)\\[0.1in]
  	=&\delta_{\alpha\beta} -4\sum_{i>j}\ Re\ (U_{\alpha i}
		U^*_{\beta i}U^*_{\alpha j}U_{\beta j})\sin^2 
		({{\Delta m^2_{ij}L}\over{4E}}),\\[0.1in]
	P_{\rm CP-odd}(\nu_\alpha \rightarrow \nu_\beta) =&-P_{\rm CP-odd}(
		\bar{\nu}_\alpha \rightarrow \bar{\nu}_\beta)\\[0.1in]
        =&2\sum_{i>j}\ Im\ (U_{\alpha i}U^*_{\beta i}U^*_{\alpha j}
          U_{\beta j})\sin ({{\Delta m^2_{ij}L}\over{2E}}),\\[0.1in]
\end{array}
\label{prob}
\end{equation}
such that,
\beq
P(\bar\nu_\alpha \to \bar\nu_\beta)= P(\nu_\beta \to \nu_\alpha) = 
P_{\rm CP-even}(\nu_\alpha \rightarrow \nu_\beta) -
P_{\rm CP-odd}(\nu_\alpha \rightarrow \nu_\beta),
\eeq
where, by CPT invariance, $P(\nu_\alpha \to \nu_\beta) = 
P(\bar\nu_\beta \to \bar\nu_\alpha)$. 
In vacuum the CP-even and CP-odd contributions are even 
and odd, respectively, under time reversal: $\alpha \leftrightarrow \beta$.

\begin{figure}[!htb]
\vspace{1.0cm}
\centerline{\epsfxsize 14.2cm \epsffile{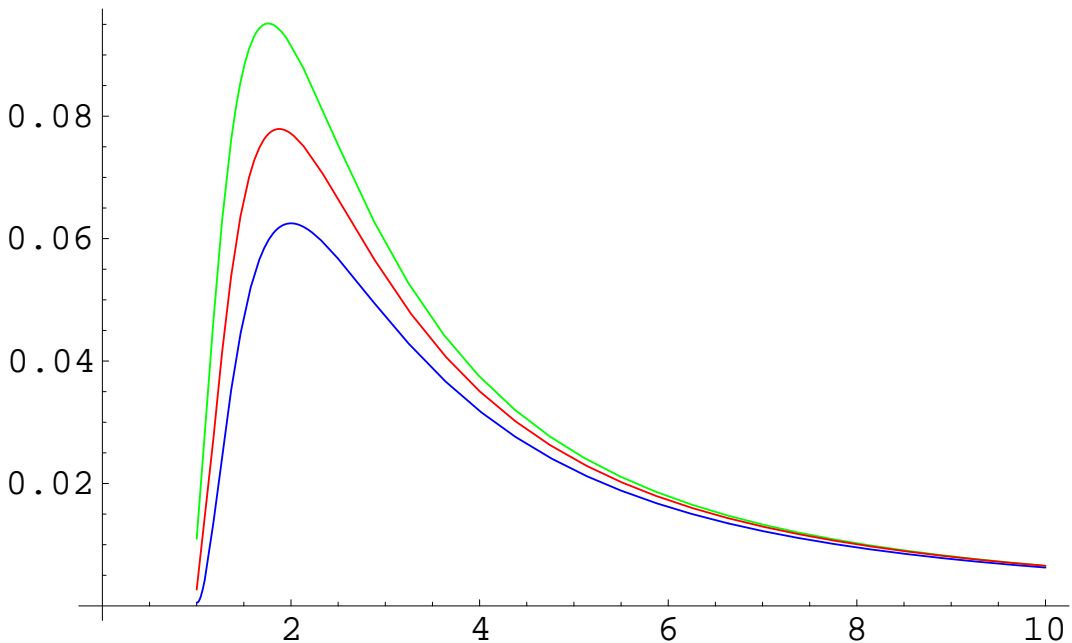}}
\caption{Transition probabilities for neutrinos (green, top curve) 
and antineutrinos 
(blue, bottom curve) in matter and vacuum  (red, middle curve) 
as function of the energy for L= 735 km,  $\Delta m^2_{13}=
3\cdot 10^{-3}~{\mbox{eV}}^2$ (normal hierarchy), 
$\theta_{\rm atm}=\pi/4$,
$\Delta m^2_\odot = 1\times 10^{-4}$~eV$^2$, $\theta_{\odot}=\pi/6$, $|U_{e3}|^2=0.04$,
and $\delta=0$. }
\label{uno}
\end{figure}

\begin{figure}[!htb]
\vspace{1.0cm}
\centerline{\epsfxsize 14.2cm \epsffile{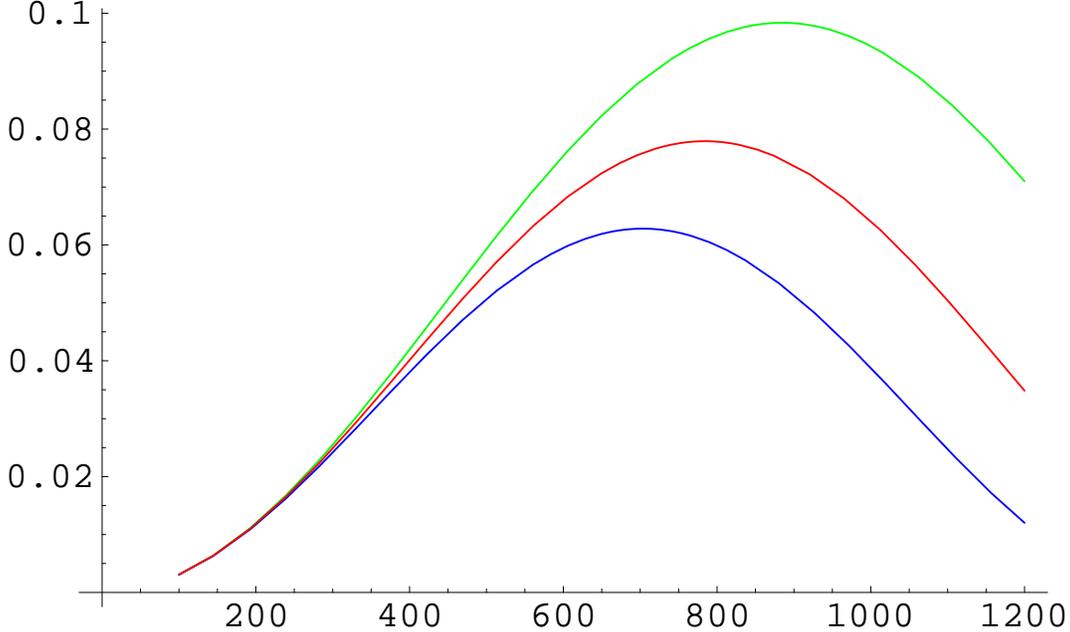}}
\caption{Transition probabilities for neutrino (green, top curve)
and antineutrinos (blue, bottom curve) in matter
and  vacuum (red, middle curve) for 2~GeV as function of the distance,
$\Delta m^2_{13}=
3\cdot 10^{-3}~{\mbox{eV}}^2$ (normal hierarchy), 
$\theta_{\rm atm}=\pi/4$,
$\Delta m^2_\odot = 1\times 10^{-4}$~eV$^2$, $\theta_{\odot}=\pi/6$, $|U_{e3}|^2=0.04$,
and $\delta=0$.}
\label{dos}
\end{figure}

If the neutrinos propagate in ordinary matter, these expressions are modified.
The propagation of neutrinos through matter is very well described by the evolution
equation
\begin{equation}
i{d\nu_\alpha\over dt} = \sum_\beta \left[ \left( \sum_j U_{\alpha j} U_{\beta
j}^* {m_j^2\over 2E_\nu} \right) + {A\over 2E_\nu} \delta_{\alpha e}
\delta_{\beta e} \right] \nu_\beta \,,  
\end{equation}
where $A/(2E_\nu)$ is the amplitude for
coherent forward charged-current scattering of $\nu_e$ on electrons,
\begin{equation}
A = 2\sqrt2 G_F N_e E_\nu = 1.52 \times 10^{-4}~{\rm eV^2} Y_e
\rho({\rm\,g/cm^3}) E({\rm\,GeV}). \,
\end{equation}
For antineutrinos, $A$ is replaced with $-A$, 
and $U$ with $U^*$.
Here $Y_e$ is the electron fraction and $\rho(t)$ is the matter density. 
For
neutrino trajectories through the earth's crust, the density is typically of
order 3~g/cm$^3$, and $Y_e \simeq 0.5$. 
For propagation through matter of constant density, the transition 
probabilities can be written in the form Eq.~(\ref{prob}) if the mass-squared
differences and mixing angles are replaced by the corresponding ``matter'' 
counterparts.
Long baseline neutrino experiments are sensitive to matter effects, and
the magnitude of the effect strongly depends on the
baseline length and neutrino energy. Some examples are depicted in 
Figs.~\ref{uno} and \ref{dos}.

There are some unknowns related to the neutrino mass pattern which can be addressed
with the ``help'' of the matter effects. As alluded to before,
the current data leave us with two alternatives for the spectrum of
the three active neutrino species: a ``normal'' neutrino mass
hierarchy or an ``inverted'' neutrino mass hierarchy.
In the case of a ``normal'' mass hierarchy, the ``solar pair'' of states is lighter
than $\nu_3$, \ie\, $m^2_3 > m^2_2, m^2_1$. In the case of inverted hierarchy, 
the states of the solar pair are heavier than  $\nu_3$, \ie\, $m^2_3 < m^2_2
\simeq  m^2_1$.
The key difference between these two hierarchies is then that, in the normal
hierarchy, the small $U_{e3}$ admixture of $\nu_e$ is in the heaviest
state whereas in the inverted hierarchy, this admixture is in the 
lightest state. 
%Then, the NMS parameterization we have presented
%above, while being ideal for the normal hierarchy case, can 
%still be used in both cases. However the identification of
%angles and processes will be different in each case.
%For example, while in the normal hierarchy $\theta_{\rm atm}$ gives
%directly the atmospheric angle, in the inverted one, the
%effective atmospheric angle will be 
%$\tan^2(\theta_{atm}) = \mid U_{\mu 1}\mid^2 / \mid U_{\mu 2}\mid^2$.
%The same happens with the angle that drives the solar oscillations,
%is given by $\theta_{12}$ in the normal case and by 
%$\tan^2(\theta_{\odot}) = \mid U_{e 3}\mid^2 / \mid U_{e 2}\mid^2 $
%in the inverted one. Notice that the CHOOZ angle in the inverted
%case is given by $ \mid U_{e 1}\mid^2$.
%While this change in angles is cumbersome and not
%particularly illuminating, there is a much simple
The difference between both schemes is parameterized by the  
sign of $\Delta m^{2}_{13}$.\footnote{Another way of treating the neutrino mass hierarchy
is by defining $m^2_3>m^2_2>m_1^2$, and redefining the solar, atmospheric and 
reactor angle depending on whether $\Delta m^2_{12}$ is larger or smaller than 
$\Delta m^2_{23}$. In such a scheme, the reactor data would limit $|U_{e3}|^2$ 
(normal hierarchy) or $|U_{e1}|^2$ (inverted hierarchy).} 
%Leaving the sign of this mass
%difference as a free parameter, has exactly the same effect
%as the change in angles we mentioned before. Now, a 
A positive (negative) $\Delta m^{2}_{13}$ will point towards a normal 
(inverted) hierarchy.
%while a negative one will do it to an inverted one.  
%This is
%the approach we will follow through this work.

\begin{figure}[!htb]
\vspace{1.0cm}
\centerline{\epsfxsize 14.2cm \epsffile{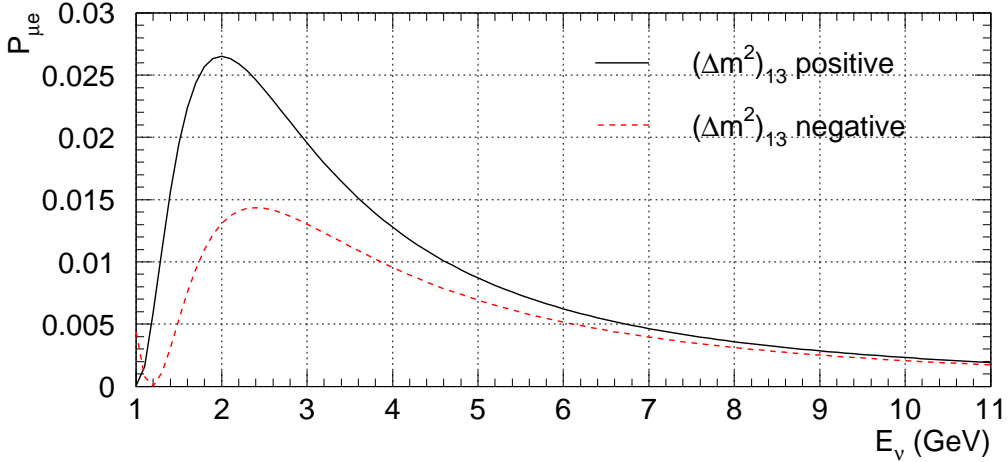}}
\caption{$\nu_{\mu}\rightarrow\nu_e$ oscillation probability for $\Delta m^{2}_{13}>0$
(solid line) and $\Delta m^{2}_{13}<0$ (dashed line). $|\Delta m^2_{13}|=
3\cdot 10^{-3}~{\mbox{eV}}^2$, 
$\theta_{\rm atm}=\pi/4$,
$\Delta m^2_\odot = 1\times 10^{-7}$~eV$^2$, $\theta_{\odot}=\pi/6$, $|U_{e3}|^2=0.01$,
and $\delta=0$.}
\label{plus_minus}
\end{figure}

In going from $\nu$ to $\overline{\nu}$, there are matter-induced CP-
and CPT- odd effects associated with the change $A \rightarrow -
A$. The additional change U $\rightarrow$ U$^*$ introduces further effects
(this is the ``genuine'' CP-violation), which are usually sub-leading. 
Note that the matter effects depend on the
interference between the different flavors and on the relative sign
between $A$ and $\Delta m^{2}_{13}$. Consequently, the experimental
distinction between the propagation of $\nu$ and $\overline{\nu}$ (the
sign of $A$) can possibly determine the sign of $\Delta m^{2}_{13}$. 
Fig.~\ref{plus_minus} shows
the transition probabilities for the two possible choices
of the sign of $\Delta m^{2}_{13}$ in the case of 
 $|\Delta m^2_{13}|=
3\cdot 10^{-3}~{\mbox{eV}}^2$, 
$\theta_{\rm atm}=\pi/4$,
$\Delta m^2_\odot = 1\times 10^{-7}$~eV$^2$, $\theta_{\odot}=\pi/6$, $|U_{e3}|^2=0.01$,
and $\delta=0$.

What is currently known about the oscillation parameters? 
The atmospheric neutrino puzzle points to a ``well determined''
mass-squared difference band \cite{SK}, 
\bea
1.5 \cdot 10^{-3}~{\mbox{eV}}^2 \lesssim \Delta m^{2}_{\rm atm}
\lesssim 5 \cdot 10^{-3}~{\mbox{eV}}^2
\eea
with $\sin^2(2\theta_{\rm atm}) \gtrsim 0.88$, while the 
solar neutrino puzzle points to many
disconnected regions. Loosely speaking, $\Delta m^2_{\odot}$
can lie anywhere between $10^{-10}$ ~eV$^2$
and $7 \times 10^{-4}$~eV$^2$ (the best upper limit is provided by
the CHOOZ experiment). Global fits to all data have historically identified
the following well-known regions \cite{solar_fits}:
\begin{itemize}
\item LMA, large mixing angle:
 $\Delta m_{\odot }^2 = (1-10)\cdot  10^{-5}$~eV$^2$, 
$\tan^2\theta_{\odot}$ = 0.2--5;
\item SMA, small mixing angle:
$\Delta m_{\odot }^2 = (4-10)\cdot  10^{-6}$~eV$^2$, 
$\tan^2\theta_\odot = (1-10)\cdot  10^{-3} $; 
\item LOW-QVO, large mixing angle, small $\Delta m^2_{\odot}$: 
 $\Delta m_{\odot }^2 = (0.5-2)\cdot  10^{-7}$~eV$^2$ 
,  $\tan^2\theta_{\odot}$ = 0.1 -- 10;
\item VO, large mixing, vacuum oscillations:  
$\Delta m_{\odot }^2 = (4-6) \cdot  10^{-10}$~eV$^2$ and
$ (6-8)\cdot  10^{-11}$~eV$^2$,  $\tan^2\theta_\odot = 0.1-3$.
\end{itemize}
Notice that all the solutions satisfy
\begin{equation}
\Delta m_{\odot}^2 \ll\Delta m_{\rm atm}^2  
\end{equation}
implying that there is indeed a hierarchy between the two distinct mass-squared 
differences.
Current global fits point towards 
electron neutrino oscillations to active neutrinos
(either muon or tau neutrinos) in the LMA region (a small 
region in the LOW area is also allowed while the 
SMA and VO regions are strongly disfavored). Although the LMA solution
is the most favored, this ambiguity must be removed by identifying
the correct solution to the solar neutrino problem. This issue will
be addressed by the KamLAND \cite{Kamland} reactor neutrino experiment,
which is capable of confirming or excluding, unambiguously, the LMA solution, 
and precisely measuring the solar oscillation parameter if LMA is indeed correct. 
However, it is important to remember that if the mass-squared difference
driving solar neutrino oscillations lies in the ``high-end'' of the LMA region,
the length of the baseline will be too long and the oscillation peaks
will not be resolved experimentally \cite{Kam_sim,Kam_sim3}. The positron spectrum
will be depleted (proportionally to the solar mixing angle) but
will not be distorted as the oscillatory term will average out.
In this case, the mass difference will be confined to a region whose
lower limit is given by the KamLAND sensitivity and whose upper limit
is dictated by the CHOOZ bound.  Thus, as this mass-squared difference is
an essential input in analyzing long baseline experiments,
the capabilities of any future determination should take this 
possibility into account. Furthermore, a solar mass difference 
belonging to the upper part of the LMA region will have a strong impact
on the oscillation probabilities, as the often neglected sub-dominant
oscillations become not that sub-dominant any more.

In order to determine the sensitivity 
of a given experiment to some parameter, it is imperative to distinguish  
different scenarios regarding the solar mass difference.
We decide to study three different scenarios:
\begin{enumerate}
\item  KamLAND does not see an oscillation signal,  implying that 
the LOW solution would be correct \cite{Kam_sim} \footnote{This is true 
only if CPT invariance holds. If CPT is broken KamLAND, using reactor 
antineutrinos, will not be able to constrain the solar (neutrino) 
spectrum \cite{nos}.}
(or, perhaps one of the disfavored non LMA solutions). In this case,  
solar driven oscillations become negligible and there is no room for
measuring CP violation in the neutrino sector. Nonetheless, one can still
attempt to determine $|U_{e3}|^2$, and, if a nonzero $|U_{e3}|^2$ is 
observed, determine the neutrino mass hierarchy. 

\item KamLAND does provide a precise measurement of the
solar mass-squared difference. The determination of the CP violating phase 
and $|U_{e3}|^2$ from a simultaneous fit to both parameters becomes possible.  

\item KamLAND sees an overall suppression of the total
rate but is not capable of measuring the mass-squared difference. In this case
one must attempt to measure not only $|U_{e3}|^2$ and $\delta$, but also the
solar mass-squared difference. While the overall picture is rather ``dirty,'' 
there is no reason to believe that a combined analysis is impossible (after all, 
for such large $\Delta m^2_{\odot}$, there will be no shortage of $\nu_e$-induced
events!).
\end{enumerate}

\setcounter{footnote}{0} 
\section{NuMI Off-Axis Beams}

The Neutrinos at the Main Injector~(NuMI)~\cite{NuMI} tertiary beamline was 
designed 
to provide an intense $\nu_{\mu}$ beam to the MINOS 
experiment \cite{MINOS}.  
The $\nu_{\mu} (\bar{\nu}_{\mu})$'s are derived from  
secondary $\pi^+ (\pi^-)$ and $K$  beams that are allowed to decay within a 
675~m decay tunnel.  
120~GeV protons will be extracted  from the  Main Injector 
via a single turn extraction ($8.6~\mu$s pulse, cycle time $1.9$~s.)
and focused downward by 58 mrad, where
the proton  impinge into a 0.94~m graphite target to produce the secondary hadron  
($\pi$, $K$) beams.  NuMI  is expected to receive 
$4\times 10^{13}$ protons/pulse.

\begin{figure}
%\centerline{\psfig{file=off_theory.eps,width=17cm}}
\centerline{\psfig{file=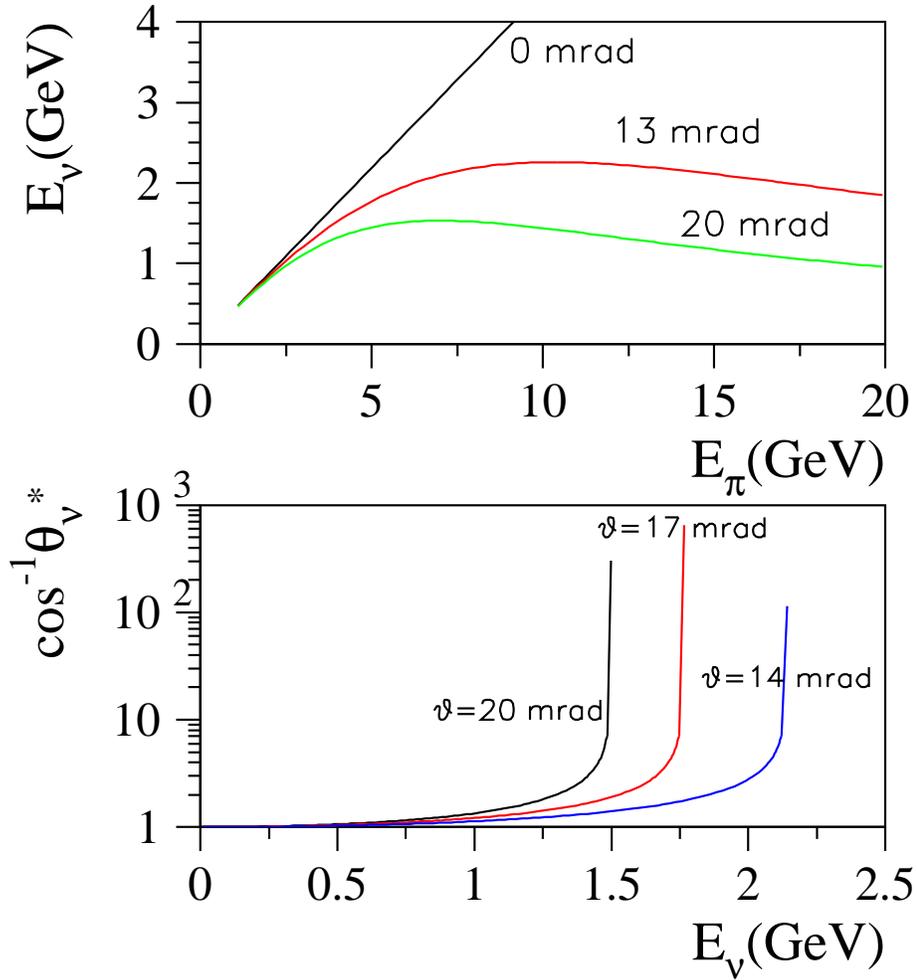,width=15cm}}
\caption[0]{\label{off_theory}
top -- allowed $E_\nu$ values for several $\theta_\nu$ values. bottom --
shows  $1/\cos\theta^*_\nu$ that is proportional to 
the neutrino flux. $\cos\theta^*_\nu=\sqrt{1-\frac{E^2_\nu}{E^{*2}_\nu}\tan^2\theta_\nu}$   is the $\nu$ production angle
in the rest frame of the $\pi$.
}
\end{figure}

MINOS is in a cavern of the Soudan mine located at a distance of 735~km from 
FNAL. The beam is tuned to  make the experiment  well aligned 
with respect to  the beam axis. Another way of efficiently using this beamline 
would be to construct detectors located  {\sl away} from the  beam axis. 
The resulting neutrino beam energy spectra at the different locations can be 
predicted from energy and momentum conservation in the $\pi$ decay process:
\begin{equation}
\label{cons}
E_\nu=\frac{m^2_\pi-m^2_\mu}{2(E_\pi-p_\pi \cos\theta_\nu)}  
     =\frac{0.004~\rm GeV}{(E_\pi-p_\pi \cos\theta_\nu)}, 
\end{equation}
where $m_\pi$ and  $m_\mu$ are the rest  masses, $E_\pi$ and $p_\pi$ are the 
pion energy and momentum, and $\cos\theta_\nu$ is the angle at which the 
neutrino is emitted with respect to the pion direction.  The  maximum angle 
in the lab frame  relative to the  pion direction is related to the 
neutrino energy by:
\begin{equation}
\label{cons_max}
\theta^{max}_\nu = \frac{(30 +\Delta_T)~\rm MeV}{ E_\nu}, 
\end{equation}
where 30~MeV is the neutrino momentum in the rest frame of the pion, 
while $\Delta_T$ takes into account the nonzero transverse momentum of the 
decaying $\pi$.
As shown in 
Fig.~\ref{off_theory}(a),  if $\theta_\nu\simeq 0$ the neutrino energy 
is proportional to the pion energy ($E_\nu=0.44E_\pi$), while at an off-axis 
location ($\theta_\nu\neq 0$) there is a maximum neutrino energy which is 
independent of the energy of the parent pion. Therefore,  the off-axis 
configuration allows one to use a fraction of the ``total'' beam that is 
characterized by having lower $E_\nu$. The maximum flux for a fixed $E_\nu$
will be obtained  when operating close to the corresponding 
$\theta^{max}_\nu$, see Fig.~\ref{off_theory}(b). The lower energy neutrinos 
provided by NuMI off-axis beams are highly desirable because they allow 
beams which are more suitable for oscillation studies, given the
current knowledge of oscillation parameters
(see Table~\ref{l_e}), while still having large enough energy and baseline length
to be sensitive to matter effects~(see Sec.~\ref{theory_sec}).

\begin{table}[t]
\begin{center}
\caption{\label{l_e} The first transition for $\nu_\mu \rightarrow \nu_X$ 
occurs when $1.27  |\Delta m^2_{13}| (L/E_\nu) = \pi/2$. Note that baselines
much greater than 900~km are not allowed in our set up, because they would require 
``levitating'' detectors.}
\bigskip
\begin{tabular}[c]{|c|c|c|c|}
\hline
\hline
$\Delta m^2_{13} (\rm eV^2)$ &  0.002      &  0.0025      & 0.003      \\
\hline
\hline
Energy (\rm GeV)             &  1.5 (2)    &  1.5 (2)     & 1.5 (2)    \\
\hline
Length (\rm km)              &  928 (1236) &  742 (989)   & 618 (824)  \\
\hline
\end{tabular}
\end{center}
\end{table}

\begin{figure}
\centerline{\psfig{file=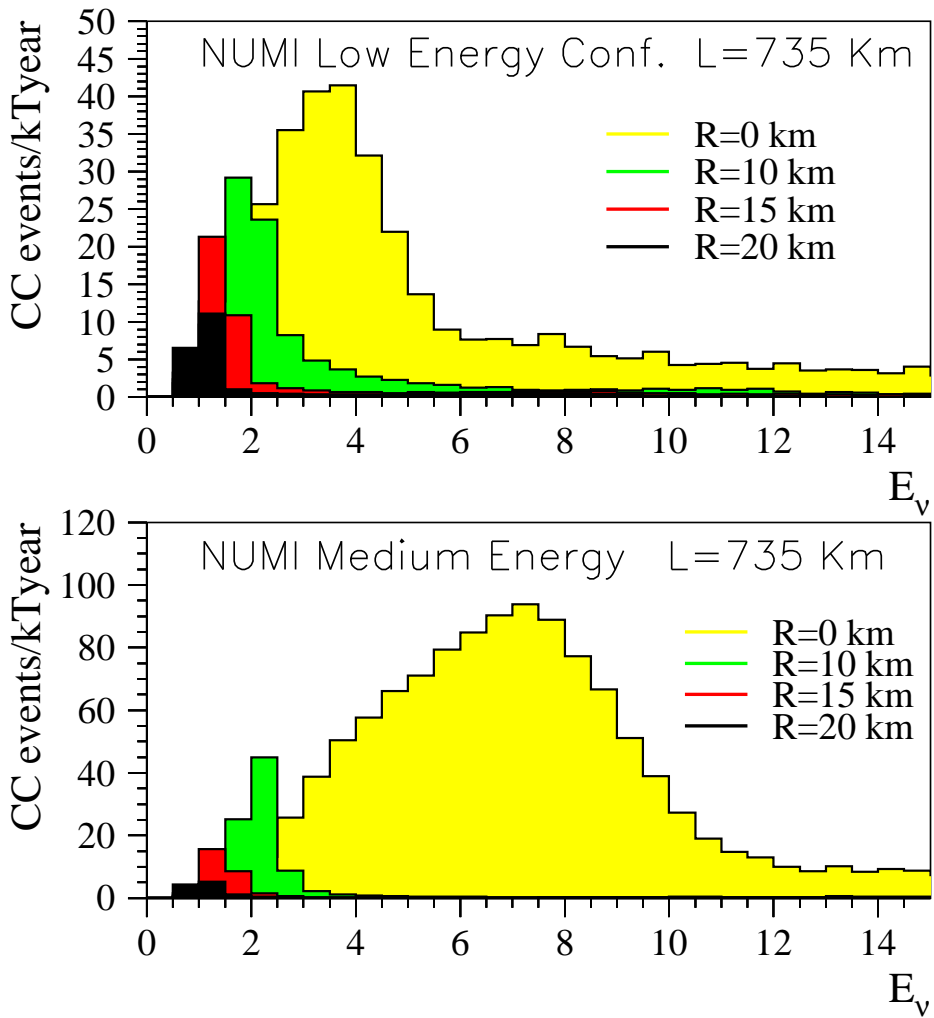,width=15cm}}
\caption[0]{\label{low_e_1} On and off-axis beams for the low and medium
energy NuMI horns configuration.  A full beam simulation  was made using the 
GEANT based GNuMI Monte Carlo. 
}
\end{figure}

Two toroidal magnetic horns sign and momentum-select the secondary 
beam.  The horns are movable allowing one to obtain different neutrino energy 
spectra.  For example, Fig.~\ref{low_e_1} depicts the expected energy 
spectrum at different location for the low energy (horns 10~m apart) and  
the medium energy (horns 27~m apart) horn configurations. 
As shown, the off-axis beams are characterized by having a 
narrow and well-defined energy distribution with fluxes   higher than the 
corresponding on-axis energy. In addition, the harmful high 
energy tail of the on-axis beams is not present, as expected from energy and 
momentum conservation, Eq.~(\ref{cons}).
All these  distributions are calculated using the GNuMI GEANT based 
Monte Carlo \cite{GNuMI}, and 
include the full beamline, target and decay pipe description.

In addition to the off-axis beam properties described in Fig.~\ref{off_theory}, 
there are two basic facts that should be kept in mind when optimizing the beam
configuration: (1) the beam flux decreases as the baseline increases, and (2)
the  beam flux decreases as the off-axis angle increases. Here we have 
performed two types of  detector location optimization.  

\begin{figure}
\centerline{\psfig{file=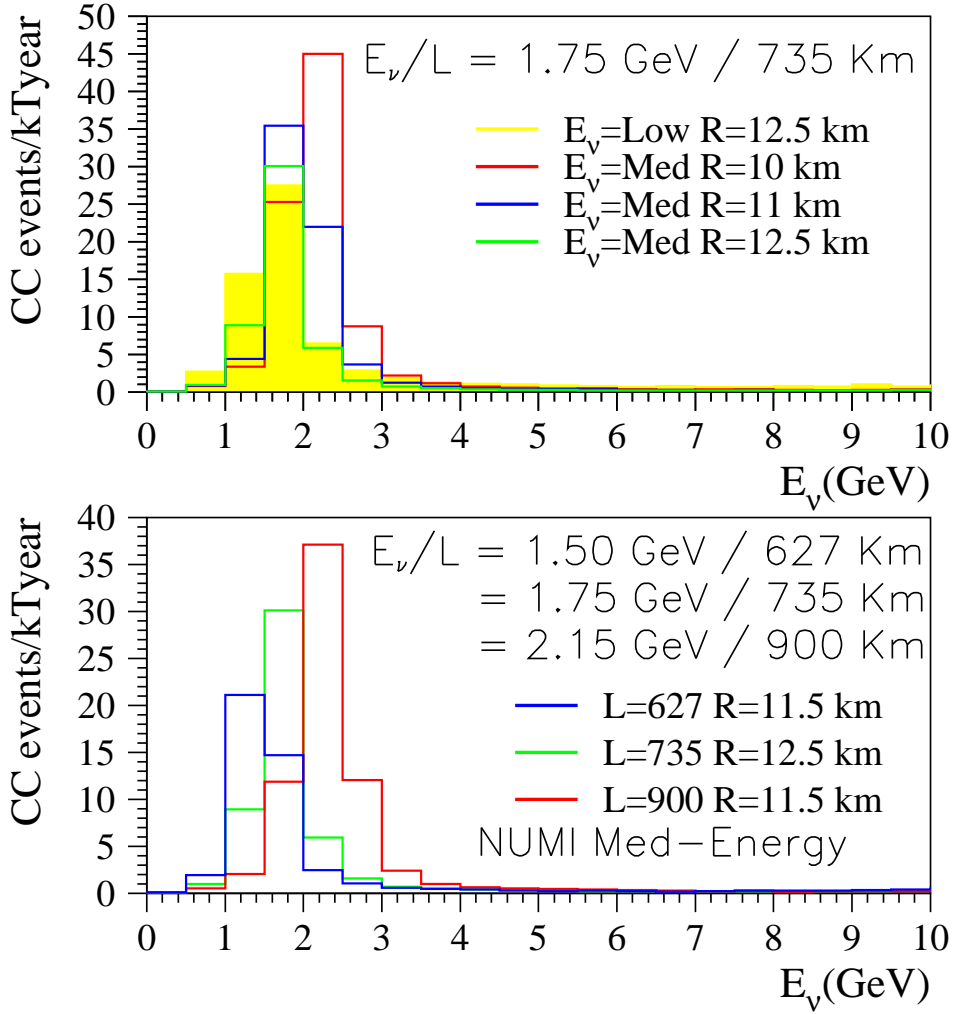,width=15cm}}
\caption[0]{\label{low_e_2}  Beam optimization for a fixed value of
$L/E_\nu$ given $\Delta m^2_{13}\simeq  0.003~\rm eV^2$. 
The medium energy 
configuration gives a higher event yield and cleaner beam conditions.
}
\end{figure}

\begin{figure}
\centerline{\psfig{file=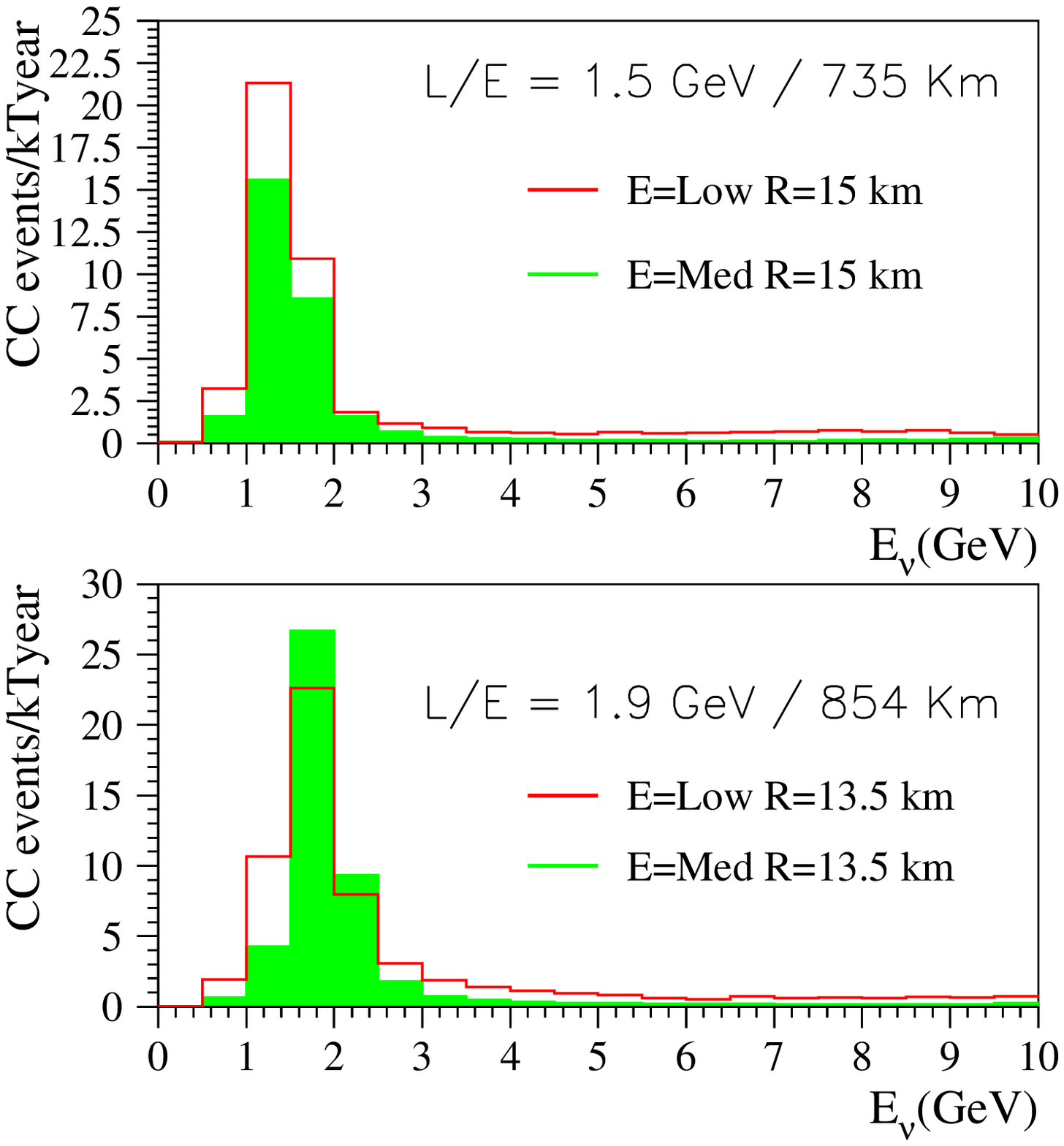,width=15cm}}
\caption[0]{\label{low_e_3}  Beam optimization for a fixed value of
$L/E_\nu$ given \mbox{$\Delta m^2_{13}\simeq  0.0025~\rm eV^2$}.
The low~(medium)  energy configuration gives a higher event yield at 
735(900)~km.
}
\end{figure}

The first one  is referred to as the naive beam optimization, where we optimize the 
detector location such that 
\mbox{$1.27 |\Delta m^2_{13}| L/E_\nu = \pi/2$}. That is, for a given 
$|\Delta m^2_{13}|$ and $L$ we have set $\theta_{exp}\simeq\theta^{max}_\nu$ 
to get the corresponding $E_\nu$ to be the maximum energy, which by definition 
will also correspond to the point of maximum flux. For example, 
$30~\rm MeV/2.25~\rm GeV \simeq 12~\rm km/900~\rm km\simeq 0.013$~mrad 
and $30~\rm MeV/1.75~\rm GeV \simeq 12.5~\rm km/735~\rm km\simeq 0.017$~mrad.
These results are depicted in Figs.~\ref{low_e_2} and \ref{low_e_3} for  
$\Delta m^2_{13}=0.003$ and 0.0025~eV$^2$, respectively.  

From this  exercise we learn that, 
if we are interested in configurations with small $\Delta m^2_{13}$, very small 
energies  will be required if the baseline is short. As a consequence,
large angles are required, say $\le 20$~mrad,  
and the lower acceptance in the medium energy beam for pions with wide angles
becomes a limitation. This can be compensated for by increasing the baseline, 
therefore increasing the required energy and lowering the required angle.
More important is the fact that, with the medium energy beam, 
there is a higher overall flux between 1.5 and 3~GeV,  and a lower high 
energy tail than with a low energy beam, if one operates 
with angles that are smaller than  $\sim$15~mrad.  These two facts make the
medium energy beam preferable.

The off-axis reduction of the high energy tail of the medium energy beam is depicted 
in  Fig.~\ref{low_e_3}. There are two reasons for this: (1) the transverse 
momentum of the 
pions is smaller for the medium energy configuration  making 
$\Delta^{medium}_T < \Delta^{low}_T$, and (2) the mean energy of the kaons
in the medium energy beam is higher producing neutrinos that are of higher 
energy and therefore  less harmful to the analysis, see Fig.~\ref{tails}.

\begin{figure}
\centerline{\psfig{file=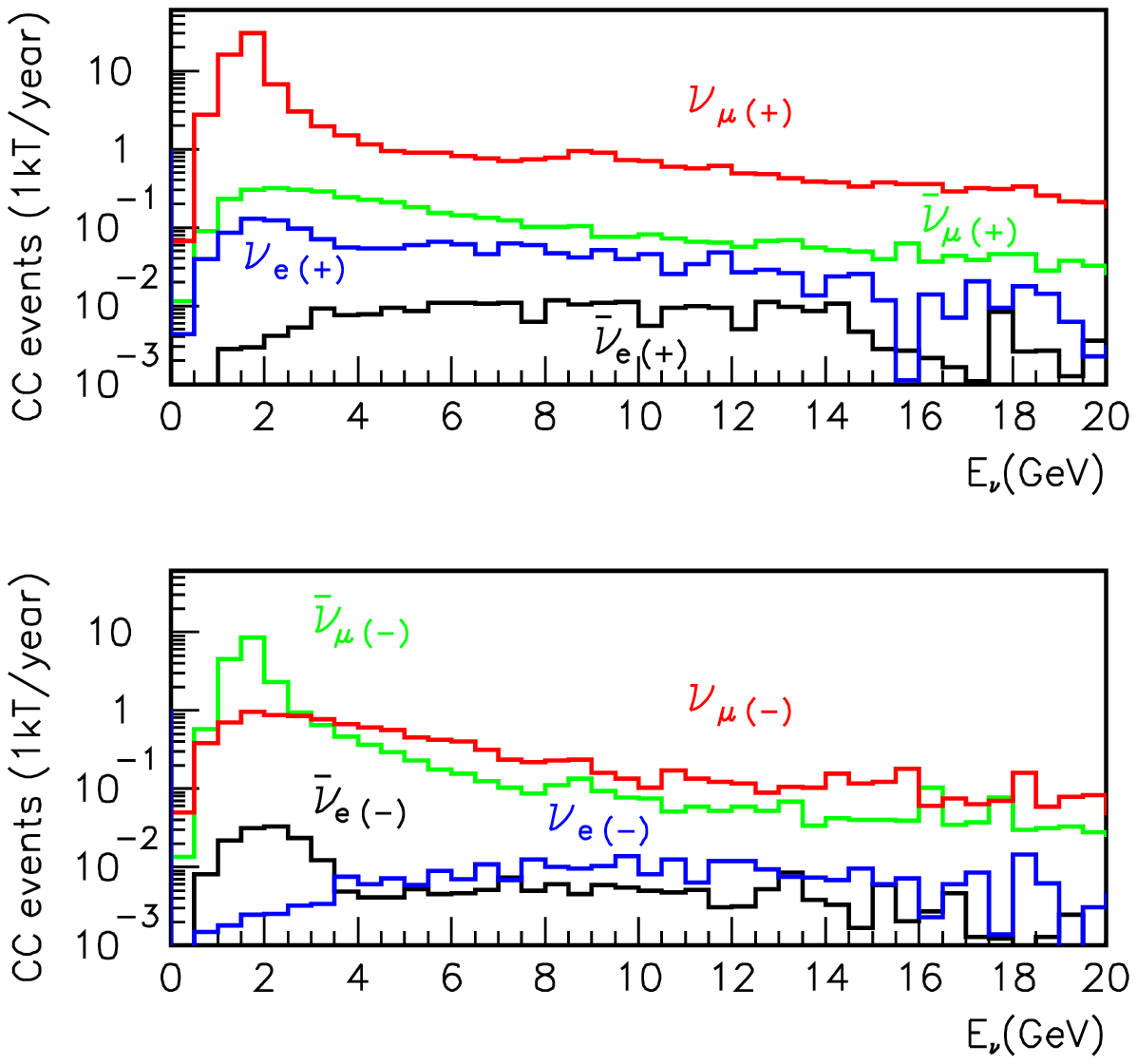,width=7.cm}
\psfig{file=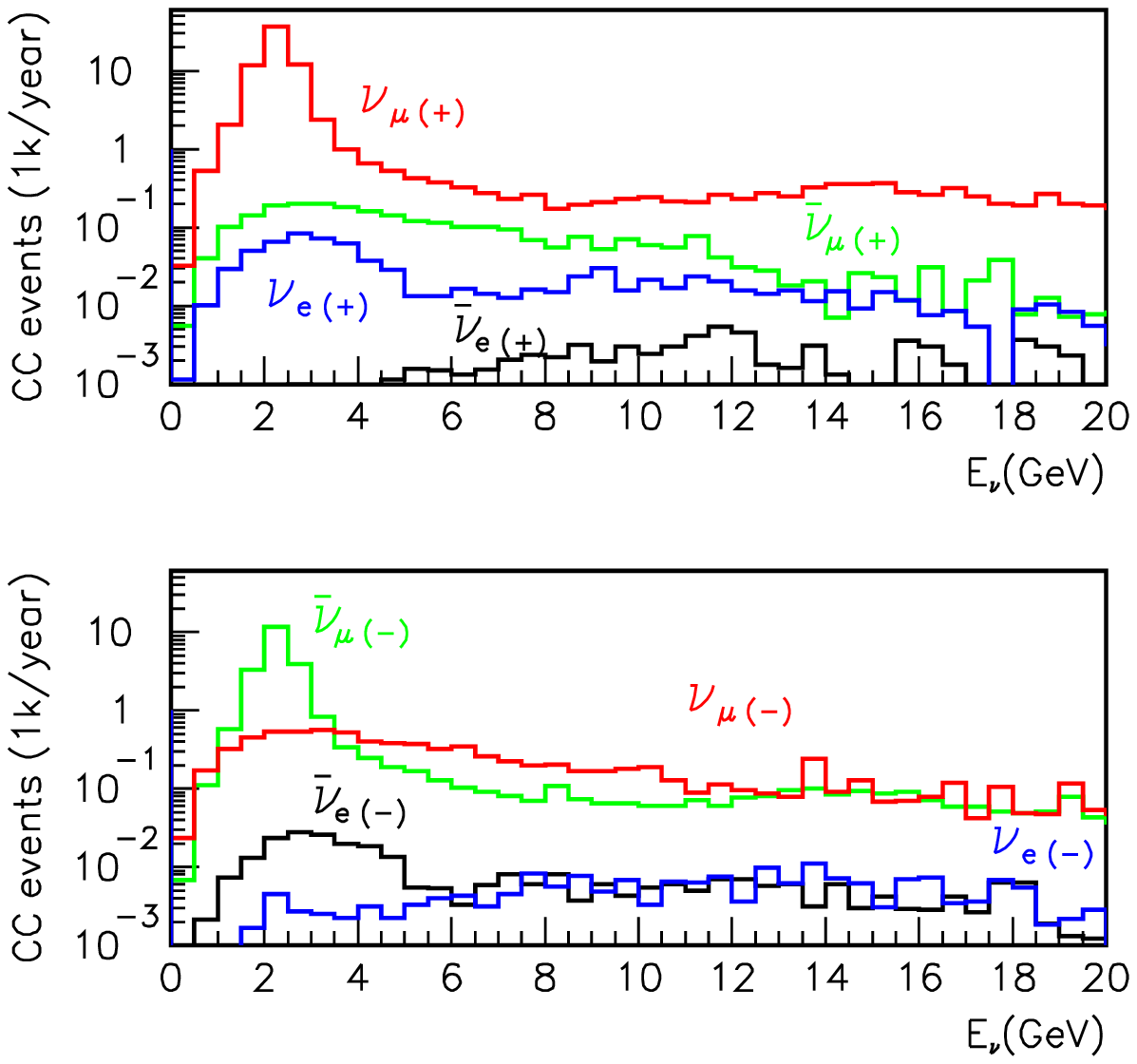,width=7.cm}}
\caption[0]{\label{beam_comp} Beam composition for positive (+)
and negative (-) horn currents. 
left: for the a low energy configuration 
at $L=735$~km and $R=10$~km.
right: for the a medium  energy configuration 
at $L=900$~km and $R=12$~km.
}
\end{figure}

\begin{figure}
\centerline{\psfig{file=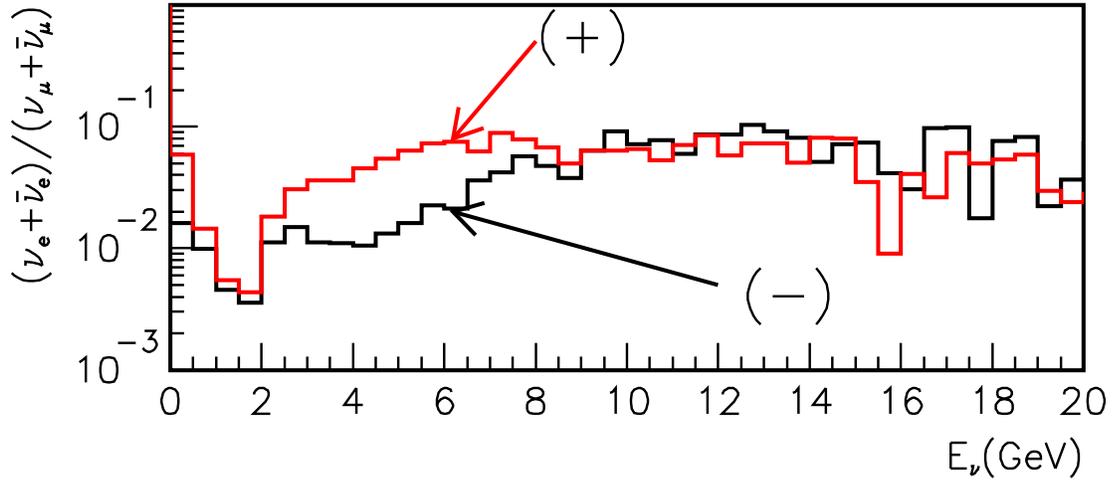       ,width=17cm}}
\centerline{\psfig{file=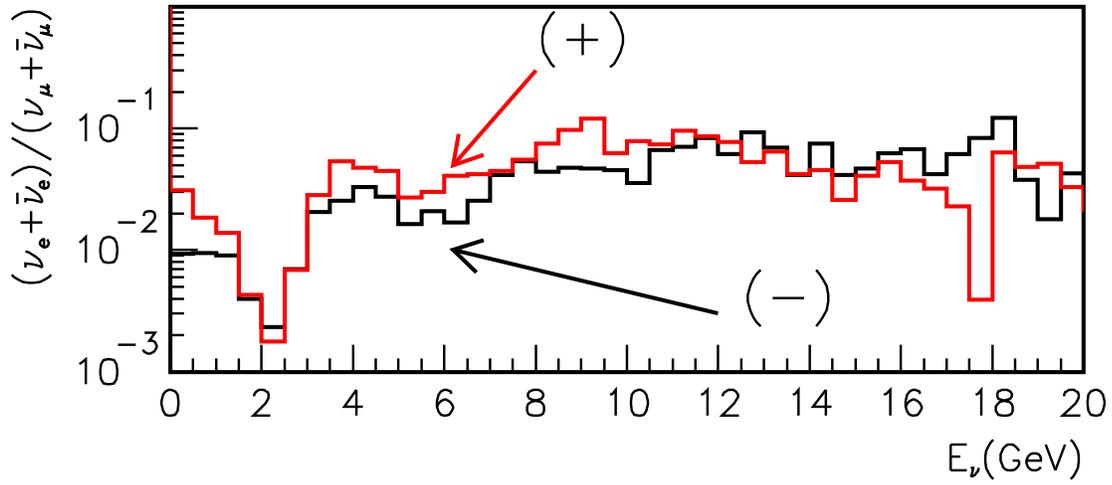       ,width=17cm}}
\caption[0]{
\label{ratio}
The $(\nu_e+\bar{\nu}_e)/(\nu_\mu+\bar{\nu}_\mu)$ does not degrade
when going from (+) to (-) polarity in the horns.  In both cases, 
the $\nu_e+\bar{\nu}_e$ fraction of the beam is below 0.5\% in the 
signal region. 
top -- is for a low energy configuration 
at $L=735$~km and $R=10$~km.
bottom: is for a medium  energy configuration 
at $L=900$~km and $R=12$~km.
}
\end{figure}

In order to determine how well different $\nu_\mu$ transitions can be 
measured, all possible backgrounds have to be taken into account.
Due to the short beam pulse duration and well defined repetition rate of the 
resulting neutrino beamline,  it is possible to  reject the cosmic ray  
background even if the detector considered  here is  
located at the surface. On the other hand, beam neutrinos induce other background
events, which must be dealt with. The flavor composition of the ``neutrino beam'' 
is depicted in Fig.~\ref{beam_comp}. 
For  the neutral current (NC) background calculations, 
all neutrino flavors contribute.
For $\nu_\mu\rightarrow\nu_e$-transitions, the inherent $\nu_e$ in the beam
contributes with further (irreducible) background events.
Assuming that we can
not detect the sign of the outgoing lepton in charged current (CC) events, the
relevant quantity is the ratio of
$(\nu_e+\bar{\nu}_e)/(\nu_\mu+\bar{\nu}_\mu)$, which 
is depicted in Fig.~\ref{ratio}.
The  $\nu_\tau$ CC cross section for $E_\nu<5$~GeV is zero~\cite{tau_cross}, 
and therefore not
relevant for the off-axis beam under consideration.

\begin{figure}
\centerline{\psfig{file=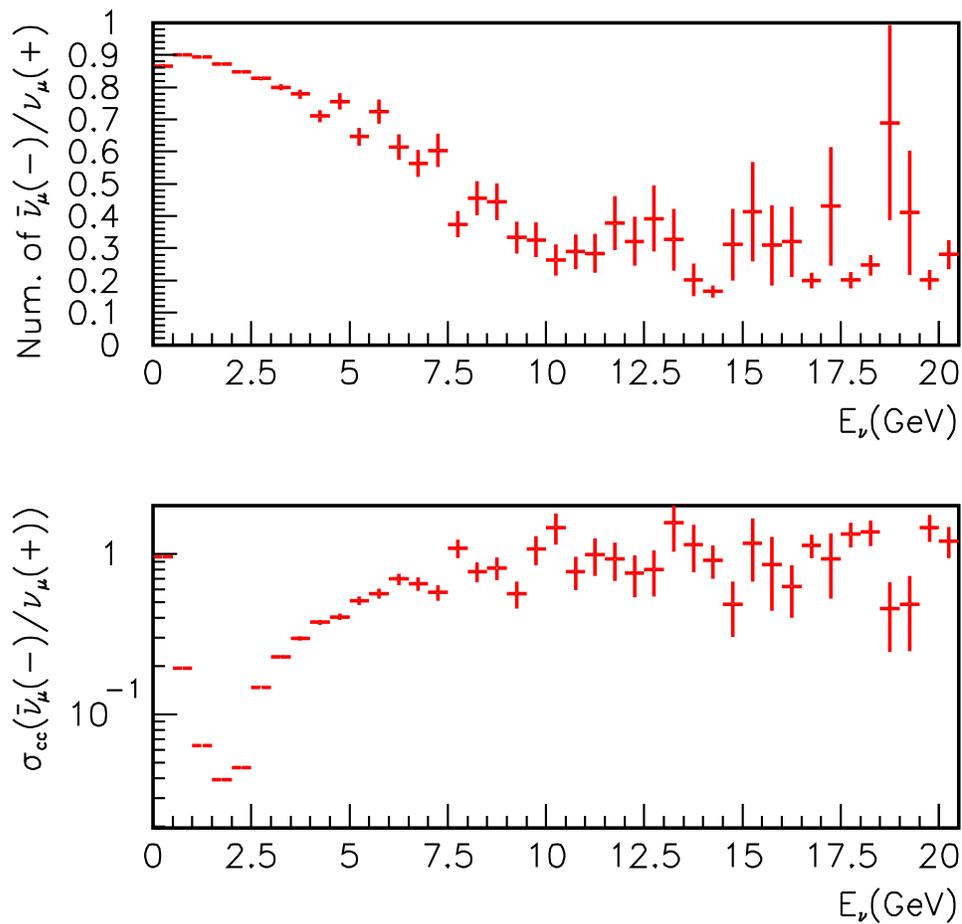,width=15cm}}
\caption[0]{
\label{cross_vs_prod} Comparison between the neutrino and anti-neutrino
beams for  charge current events.  As shown, the lower rate is due 
to the cross section and not to the $\pi$  production rate.
}
\end{figure}

Studies of matter effects and CP violation require the comparison of
$\nu_\mu$ and  $\bar{\nu}_\mu$ oscillations. An ``anti-neutrino beam'' can be
produced by reversing the polarity of the horns.  
This beam's flavor composition and $(\nu_e+\bar{\nu}_e)/(\nu_\mu+\bar{\nu}_\mu)$ 
are also depicted in Figs.~\ref{beam_comp} and
\ref{ratio}. The beam related background is  at the same level for
both horn polarities, but the CC event rate is at least three 
times smaller for $\bar{\nu}_\mu$  with energies above 2.5~~GeV and 
five times smaller for the lower neutrino energies. As shown in 
Fig.~\ref{cross_vs_prod}, this is mostly caused by the cross section
difference, and not by the $\pi^+$, $\pi^-$ production rate.
 
\begin{figure}
\centerline{\psfig{file=pion_kaon.epsi,width=14cm}}
\caption[0]{\label{tails}  Contribution  from  $\pi$ and kaons to the
 beam composition for off-axis beams in the  medium and low energy beam 
horn configuration.
}
\end{figure}
  
\begin{figure}
\centerline{\psfig{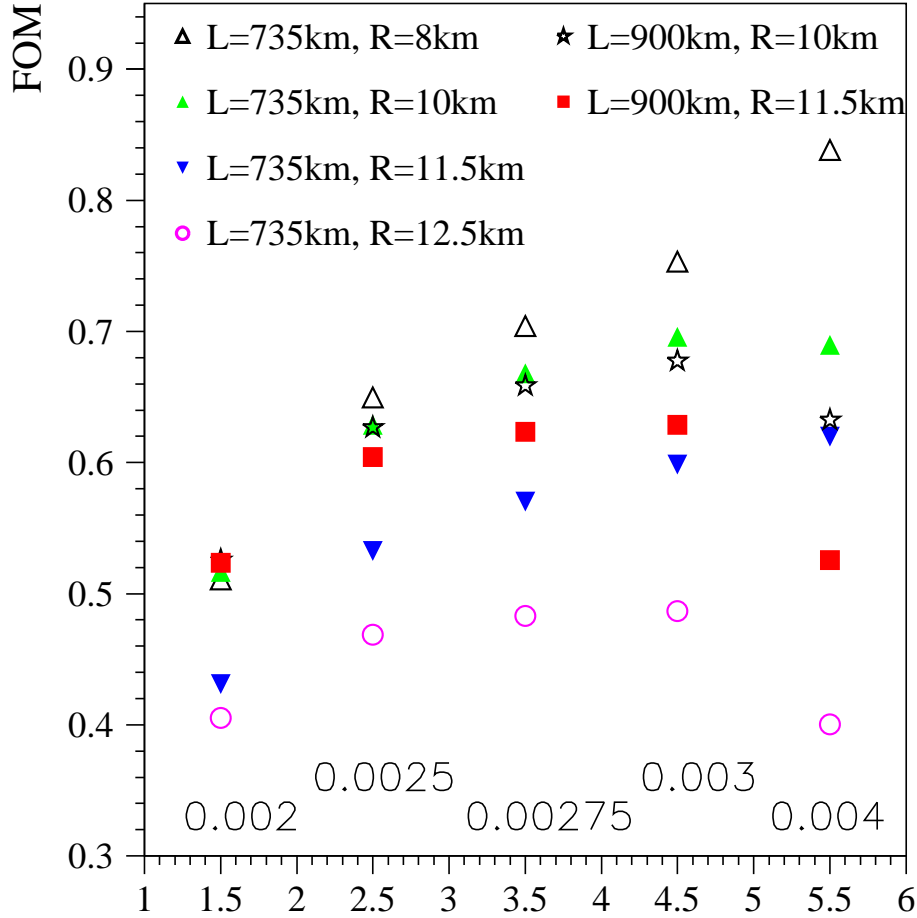}}
\caption[0]{
\label{fom}FOM for
$\nu_\mu \to \nu_e$ appearance for different values of $\Delta m^2_{13}$  
in the case of full mixing and $|U_{e3}|^2=0.01$ for different baselines 
and $\theta_\nu$, but with the same detector and
reconstruction criteria.  We only select events with visible energy between
1 and 3~GeV. If this highly segmented detector were 
at the MINOS location  
 and $\Delta m^2_{13}$ were $0.003$~eV$^2$, the corresponding FOM
would be 0.54.  
This is to be compared to the FOM for the MINOS detector that 
is only 0.39~\cite{milind}. This means, that if we ignore
the background uncertainty,  about half of the
gain in sensitivity is due to the location and the other half to the detector.}
\end{figure}

\begin{figure}
\centerline{\psfig{file=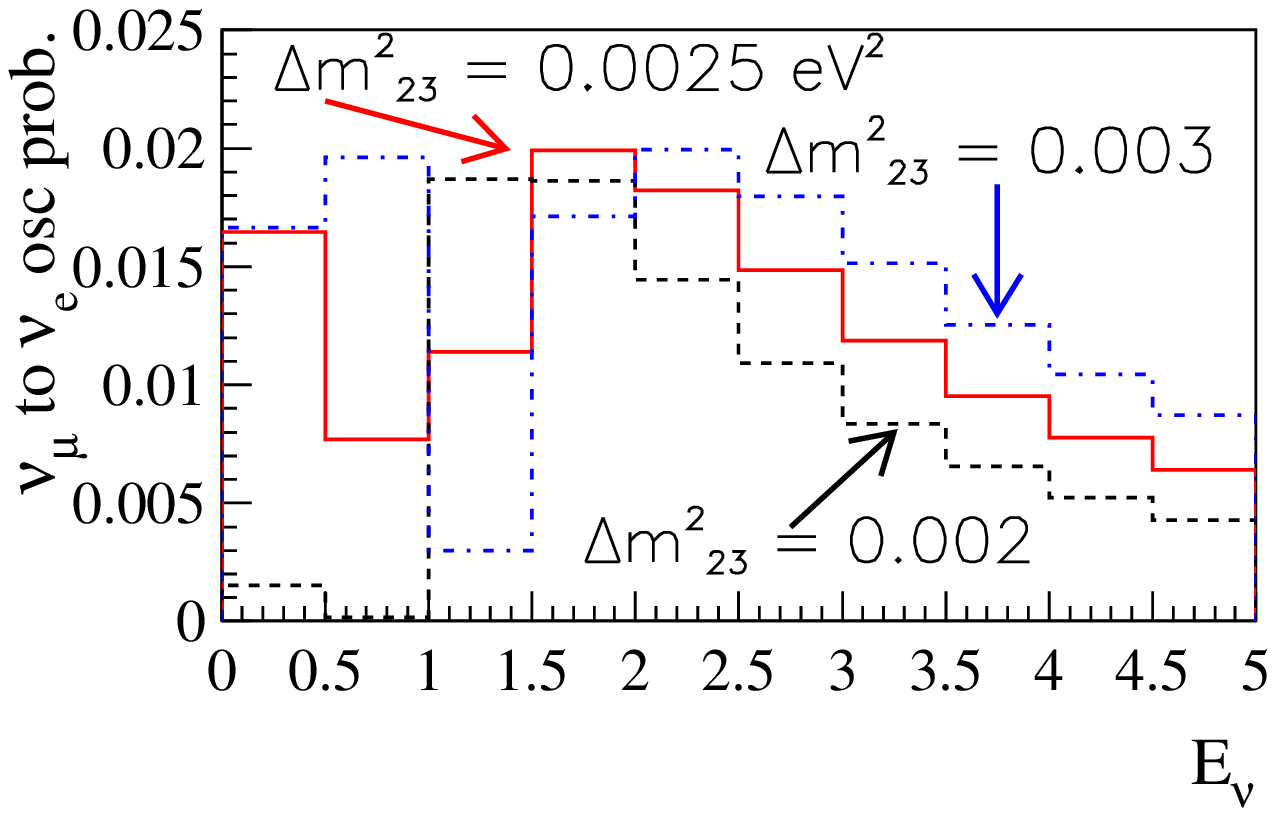,width=12cm}}
\caption[0]{
\label{prob1} 
Transition probability for $\nu_\mu \to \nu_e$ in the case of maximal atmospheric mixing and
$|U_{e3}|^2=0.01$, for different values of $\Delta m^2_{23}$. We assume $\Delta m^2_{12}\ll
10^{-5}$~eV$^2$.}
\end{figure}

In the second type of beam optimization, we look at locations 
with smaller $ \theta_\nu$.  These locations are characterized by having a  
broader energy spectrum  and    a
higher overall flux for neutrinos with energies below 3~GeV. This is 
illustrated in 
Fig.~\ref{low_e_2}(top), where we show the energy distributions at   several
radii and at a fixed length.  Clearly, for a detector located at a radius of
10~km~(13~mrad) instead of 12.5~km~(17~mrad), we have at least two times more 
flux between 1.5 and 3~GeV.  These  broader energy beams are suitable 
for a wider range of  $\Delta m^2_{13}$, but how useful they are for 
$\nu_\mu \to \nu_e$ appearance experiments will depend on how 
well one can keep the NC background under control. 
Therefore, in order to conclude how much better these larger 
$ \theta_\nu$ off-axis 
experiments really are requires a full evaluation of 
the signal and backgrounds with a realistic detector simulation and 
reconstruction.  We have  performed such analysis and the details are 
given in Sec.~\ref{rec_section}. Here we will just summarize the results
by evaluating a figure-of-merit~(FOM) at each location for
$\nu_\mu \to \nu_e$ in the case of full mixing and $|U_{e3}|^2=0.01$.
We defined the FOM as $S/\sqrt{S+B}$, where  
$S$ and $B$ are the signal and background events 
that survive all the cuts in the reconstruction. As depicted in Fig.~\ref{fom},
off-axis experiments with angles between  $10  \le \theta_\nu \le 13$~mrad 
from the axis have a  high FOM for all values of $\Delta m^2_{13}$.  
The high FOM is  not only due to the characteristics of the beam and 
oscillation probabilities (see Fig.~\ref{prob1}), but also due to the fact 
that in all cases we can keep the NC background at the 0.5\% level,
while the reconstruction efficiency is about 40\%.
If we look in detail at the case of 
$\Delta m^2_{13} = 0.003~\rm eV^2$, we can see that the naive beam tune performed
at 735~km cannot compete with smaller $\theta_\nu$ locations at the
same baseline. This is not true for the naive beam tune performed 
at 900~km, where at that location $\theta_\nu$
is already small enough to give us a high integrated flux.  We can still obtain
a 20\% increase in the FOM by reducing the baseline to 735~km
and $\theta_\nu$ to 10~mrad, but at the moment this is not particularly relevant,
given the current lack of knowledge of neutrino oscillations parameters. For example,
in Fig.~\ref{prob1} we have assumed a normal neutrino mass hierarchy, and an
inverted hierarchy would significantly modify what the optimal conditions are. 
Furthermore, the inclusion of (potentially significant) solar 
oscillations and CP-violating effects further cloud the picture.
%we have to many uncertainties
%in the mixing parameters to 
%decided if this reduces our sensitivity to matter effects.
Nonetheless, the FOM is flat enough that any ``reasonable'' choice of baseline
and opening angle should be ``close'' to optimal. 
We, therefore,  will do all our analyses for a baseline of  
900~km and at a radius of 11.5~km.

 \subsection{NuMI Off-Axis Beams With A New Proton Driver}
 The Proton Driver design described in~\cite{pdriver, booster} will allow us 
 to bring the NuMI neutrino beam power up from 0.4~MW to 1.6~MW. 
 This design  is based on an 8~GeV circular machine with a circumference  
 of 473.2~m, and it will provide $2\times 10^{13}$ protons per 
pulse instead of the assumed $5\times 10^{12}$
of the current booster. 
 In addition, the total luminosity could be further increased by 30\% if
 the current linac gets a 200~MeV upgrade. In this case,  we would get 
 $3\times 10^{13}$  protons per pulse.  This machine is estimated to cost
 US\$160M.
 
An  alternative design made out of only a  linac to accelerate 
protons up to  8~GeV,
using SNS and Tesla style superconducting cavities, is  also under 
consideration~\cite{foster}.

\section{Detector Simulation and Expected Signal Efficiencies}
\label{rec_section}

To determine our $\nu_\mu\to\nu_e$ detection capabilities, 
we are considering a highly segmented iron-scintillator detector.
A full study of the expected  detector performance, event reconstruction 
efficiencies, and background contamination was performed.

The detector is made up of 4.5\,mm thick iron foils (one quarter of one
radiation length) interleaved with 1 cm thick scintillator planes.
The scintillator strips are oriented at plus or minus $45^\circ$
from vertical, alternating every other plane.  The width of each
readout cell is 2\,cm.  A 3\,cm long air gap is left between each two
iron-scintillator pairs to further improve the detector performance
by increasing the effective radiation length at no additional cost.
This choice is a compromise between the need for a good separation of
two close electromagnetic showers on the one hand, and good clustering of
each individual shower on the other.

Signatures of $\nu_e$ and $\nu_{\mu}$ CC events, as well as of NC
events, were studied in GEANT-based Monte Carlo simulations, with
the GMINOS program developed at Fermilab~\cite{R_Hatcher}.  
Typically, a 1--2 GeV $\nu_e$ induced shower will
leave hits in 10-20 consecutive planes while $\nu_\mu$ ones
will do it in around 40 consecutive planes, making possible a complete
track finding procedure.  High transverse segmentation provides good
separation of two close tracks; preliminary studies showed that
approximately 66\% of $\pi^0$'s at 1-2 GeV can be successfully
identified as producing two well separated showers in at least one
view.

The event reconstruction consists of track fitting and track selection.
Tracks are fitted and examined in each view separately.  A good track
is required to give hits in at least 4 planes and have good $\chi^2$
for a straight line.  Most tracks coming from charged pions and
recoiling protons can be rejected by requiring a mean track width
of at least two cells and a width at maximum of at least three cells.
``Baby tracks", that is, tracks found in the vicinity of the end of
a longer main track and less than half of its length, come mostly
from secondary particles within the same shower and are discarded.
Conversely, two tracks of comparable length and/or pointing at the
same interaction vertex are a signature of a NC event with a $\pi^0$
in the final state.

After track selection, a signal candidate event is required to leave
exactly one good track in each view.  Additional selection criteria
are imposed on the event basis.  To optimize the ratio of $\nu_e$
signal to intrinsic $\nu_e$ background, a window in the total visible
energy is defined (for $\Delta m^2_{\rm atm} = 0.0025 - 0.003$ GeV$^2$ a
reasonable choice is 1-3 GeV).  A small missing $p_T$ with respect to
the beam direction is required, a minimum fraction of the total event
energy carried by the track (both criteria helping reject NC), and
no track longer than 28 planes in any view (suggestive of a muon).
The remaining NC background is further reduced by checking for a
displacement of the beginning of the shower with respect to the
interaction vertex, the latter being identified by the trace of
the recoiling proton (if any).

\begin{figure}
\centerline{\psfig{file=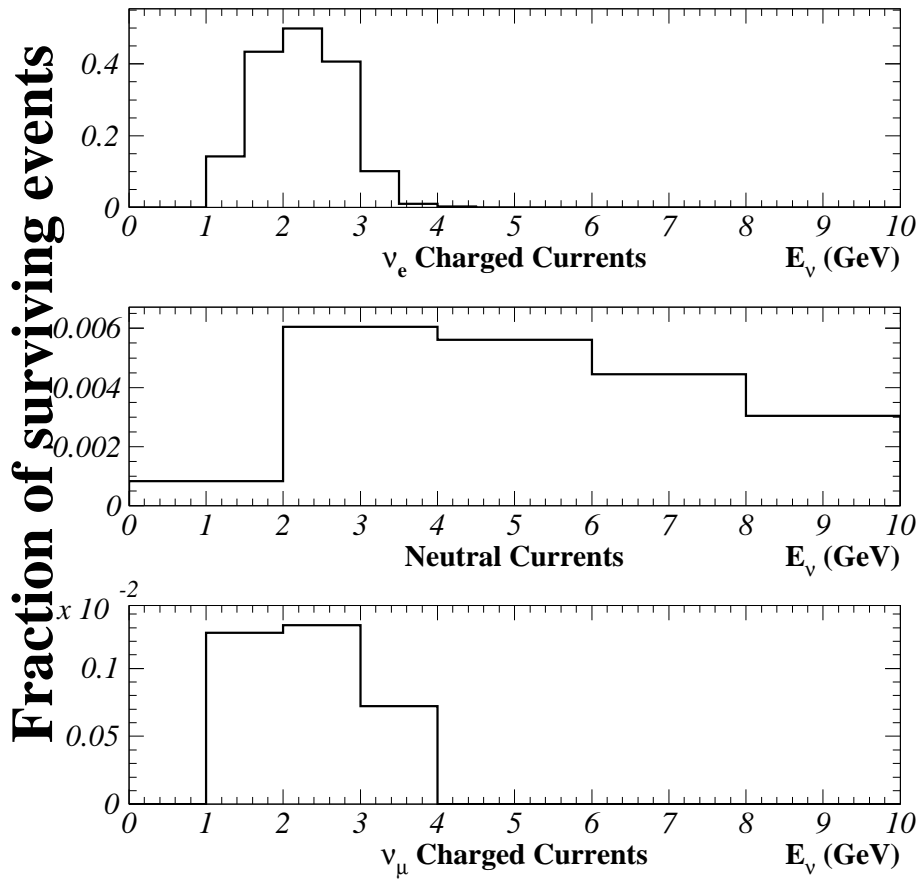,width=12cm}}
\caption[0]{
\label{fig_rec} Reconstruction efficiency for the selection of  $\nu_e$ CC
events, the NC background and the $\nu_\mu$ CC events for 
highly-segmented iron scintillator calorimeter.
}
\end{figure}

The above criteria allow a rejection of about 99.7\% of all NC events,
while $\nu_{\mu}$ CC are suppressed to a negligible level.  The
resulting reconstruction efficiency and remaining contaminations from
NC and $\nu_{\mu}$ CC as a function of the incoming neutrino energy
are shown in Fig.~\ref{fig_rec}.

In order to evaluate how well we can do $\nu_\mu\to\nu_e$ searches
in the different off-axis beams locations, we have convoluted the beam
spectra shown in Fig.~\ref{low_e_2} and the resulting 
reconstruction efficiencies shown in 
Fig.~\ref{fig_rec}.  The results are summarized in 
Table~\ref{rec_tab} and Fig.~\ref{fom}. 

\begin{table}[t]
\begin{center}
\caption{\label{rec_tab} Expected event yields  in the case of  
$\nu_\mu\to\nu_e$ appearance for 1~kton-year  for a detector
located at different baselines (L) and distance from the beam axis (R), 
and different horn polarity ($\pm$).  We assume that we are in vacuum, 
and maximal mixing solutions with $\Delta m^2_{13}=0.003~\rm eV^2$ and 
$|U_{e3}|^2=0.01$.
}
\bigskip
\begin{tabular}[c]{|c|c|c|}
 \hline \hline
L, R, $\pm$ & {735~km  0~km  +}& { }\\ \hline
Signal CC $\nu_e$  & 0.709/3.351 = 0.212 & \\ \hline
CC $\nu_e$ & 0.297/6.339 = 0.047 & \\ \hline
CC $\nu_\mu$  &  0.053/378.391 = 0.0001 &\\ \hline
CC $\nu_\tau$  &  0.059/4.505 = 0.013 &\\ \hline
NC  &  0.613/165.686 = 0.0037  & \\ \hline
\hline
L, R, $\pm$ & {735km 10km  +} &{735km 12.5km  + }\\ \hline
Signal CC $\nu_e$ & 0.687/1.578    = 0.435  & 0.342/0.897 = 0.381 \\ \hline
CC $\nu_e$  &  0.145/1.413     = 0.102 & 0.105/0.825     = 0.127 \\ \hline
CC $\nu_\mu$  &  0.012/26.97   = 0.0005  & 0.0062/15.554= 0.0004\\ \hline
NC   & 0.132/32.1816  = 0.0041 & 0.0411/18.363 = 0.0022\\ \hline
\hline
L, R, $\pm$ & {900~km  11.5~km  +}& {900~km   11.5~km   - }\\ \hline
Signal CC $\nu_e$  & 0.554/1.282 = 0.432 & 0.183/0.460 = 0.399\\ \hline
CC $\nu_e$ & 0.10145/1.036    = 0.098 & 0.038/0.561     = 0.067\\ \hline
CC $\nu_\mu$  & 0.0069/18.04 = 0.0004&0.0027/10.924= 0.0002\\ \hline
NC   & 0.113/24.950  = 0.0046&0.045/10.346 = 0.0044\\ \hline
\hline
\end{tabular}
\end{center}
\end{table}

\section{Simulated Data Analysis}

\subsection{Not the LMA Solution}

If KamLAND does not observe a suppression of the reactor antineutrino 
flux, the LMA solution to the solar neutrino puzzle will be excluded 
\cite{Kam,Kam_sim}, indicating that $\Delta m^2_{12}\ll 10^{-5}$~eV$^2$ and/or 
$\tan^2\theta_{\odot}\ll 1$. In this case, it is well known that
the CP-odd phase $\delta$ is not observable in standard long-baseline
experiments, not only because solar oscillation do not have enough time
to ``turn on,'' but also because matter effects effectively prohibit
any neutrino transition governed by the solar mass-squared difference.
This being the case, one can only study $\nu_{\alpha}\leftrightarrow\nu_{\beta}$
transitions governed by the atmospheric mass-squared difference. For the
reasons outlined earlier, we will concentrate on $\nu_{\mu}\rightarrow\nu_e$
transitions and, possibly, $\bar{\nu}_{\mu}\rightarrow\bar{\nu}_e$. 

As mentioned in the Sec.~3, for the same running time, 
the number of interactions due to muon-type neutrinos
obtained with the ``neutrino beam'' is significantly larger than the
number of interactions due to muon-type antineutrinos obtained with the 
``antineutrino beam.'' Therefore, it seems logical to start running with 
the ``neutrino beam'' and decide whether one can observe an excess of 
$\nu_e$-like events
in the off-axis detector. This is done by simulating data and
testing whether the observed number of events is significantly 
more than the the expected number of background events. We define the
$\chi$-sigma average sensitivity region by
\begin{equation}
\chi^2=\frac{(DATA-BKG)^2}{DATA+\sigma^2_{BKG}}+1
\end{equation}
where $DATA$ is the averaged number of observed events for a given
set of theoretical parameters, while $BKG$ is the expected number of
background events. $\sigma_{BKG}$ is the expected error on the 
estimation of the number of background events. Unless otherwise
noted $\sigma^2_{BKG}=(0.1\times BKG)^2$. In practice
$DATA=SIGNAL(U_{e3},\Delta m^2,\delta,\cdots)+BKG$. The ``$+1$''
accounts for the fact that we are computing the sensitivity of an
average experiment (see \cite{seasonal} for details).   

\begin{figure}
\centerline{\epsfxsize 14.2cm \epsffile{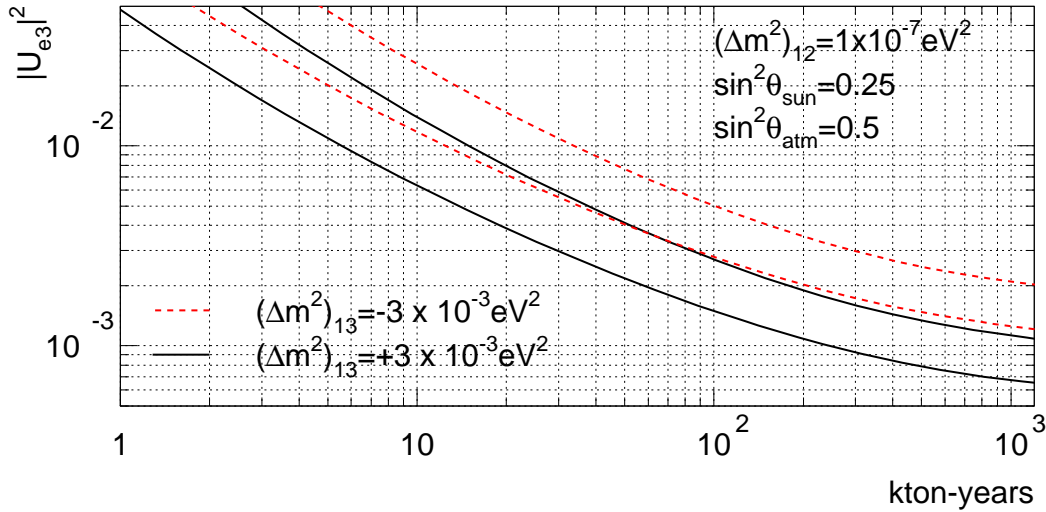}}
\caption{Two and three sigma sensitivity reach for $|U_{e3}|^2$ as a function of the
running time (in kton-years), for a normal neutrino hierarchy (solid lines) and an inverted
neutrino hierarchy (dashed lines). $|\Delta m^2_{13}|= 3\times 
10^{-3}$~eV$^2$, $\sin^2\theta_{\rm atm}=1/2$ and $\Delta m^2_{12}=
10^{-7}$~eV$^2$, $\sin^2\theta_{\odot}=1/4$, $\delta=0$.
For a normal neutrino mass hierarchy, a 20~kton detector with (without) an
proton driver upgrade will be able to detect a three sigma signal for
$\nu_\mu\to\nu_e$ oscillations, if
$|U_{e3}|^2\le 0.0015(0.0028)$ after 5 years.}
\label{kton-year}
\end{figure}

Fig.~\ref{kton-year} depicts the two and three sigma
sensitivity to $|U_{e3}|^2$ as a
function of the number of kton-years of accumulated neutrino beam   
data collected off-axis, in the case $\Delta m^2_{13}=\pm 3\times 
10^{-3}$~eV$^2$, $\sin^2\theta_{\rm atm}=1/2$ and $\Delta m^2_{12}=
10^{-7}$~eV$^2$, $\sin^2\theta_{\odot}=1/4$ (for concreteness. The precise value
of the solar parameters is irrelevant in the case at hand). 
The definition of one kton-year is the following: it is the amount of
events observed after one year of running of the current NuMI beam (see
Fig.~\ref{beam_comp}) in a one kton detector at 900~km, 11.5~km off-axis.
Many features are
readily noticeable. First, in order to be sensitive to values of
$|U_{e3}|^2$ which are significantly smaller than the current 
CHOOZ bound ($|U_{e3}|^2\lesssim 0.05$ \cite{CHOOZ}), one is required to 
accumulate more than 40~kton-years of data, in the case of a normal hierarchy,
or more than 150~kton-years in the case of an inverted hierarchy.
This means, assuming the nominal NuMI beam, roughly two or eight 
years of running with a 20~kton detector.  
Note that the $10\%$ uncertainty 
on the background determination dictates that, even after accumulating
an infinite amount of statistics, the three sigma 
reach of the off axis experiment plateaus at around $|U_{e3}|^2\sim 1\times 
10^{-3}$ ($2\times 10^{-3}$) for a normal (inverted) hierarchy.
As is clear from Fig.~\ref{kton-year}, 
in the case of an inverted hierarchy, the sensitivity is
significantly worse. This is also expected, since matter effects enhance
the $\nu_e$ appearance rate in the case of a normal hierarchy and
reduce it in the case of an inverted hierarchy (see Fig.~\ref{plus_minus}).

Note that the sensitivity would be significantly different 
for different values of $|\Delta m^2_{13}|$ and that, by design, the 
sensitivity is optimal at around $|\Delta m^2_{13}|\sim 3\times 
10^{-3}$~eV$^2$. We have verified, as discussed earlier, that it 
does not deteriorate significantly for 
$|\Delta m^2_{13}|\sim (2-4)\times 10^{-3}$~eV$^2$.

\begin{figure}
\centerline{\epsfxsize 13cm \epsffile{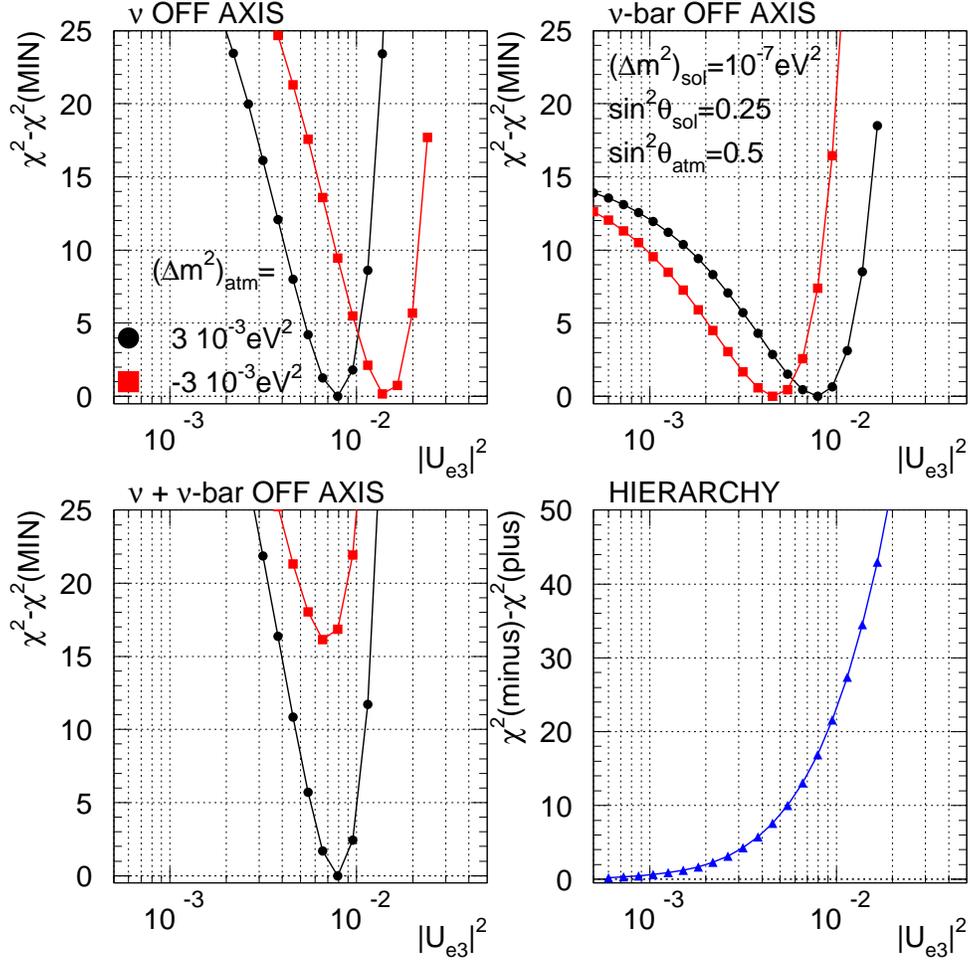}}
\caption{top-($\chi^2-\chi^2_{\rm min}$) as a function of $|U_{e3}|^2$, assuming both neutrino
mass hierarchies, upon analyzing simulated data consistent with $\Delta m^2_{13}>0$ and
$|U_{e3}|^2=0.008$ after 120 kton-years of neutrino-beam running (left) and 300 kton-years
of antineutrino-beam running (right). bottom,left-same as above, after combining the
two data sets. bottom,left-difference of minimum value of $\chi^2$ obtained with the hypothesis
$\Delta m^2_{13}>0$ and $\Delta m^2_{13}<0$ as a function of the input $|U_{e3}|^2$. See text
for details. $|\Delta m^2_{13}|= 3\times 
10^{-3}$~eV$^2$, $\sin^2\theta_{\rm atm}=1/2$ and $\Delta m^2_{12}=
10^{-7}$~eV$^2$, $\sin^2\theta_{\odot}=1/4$, $\delta=0$.}
\label{measure_low}
\end{figure}

If one detects an excess of $\nu_e$-like events, the next step is to 
determine the value of $|U_{e3}|^2$. Again, we do this by performing
a $\chi^2$ fit to the ``data.'' We will assume that the atmospheric
parameters $|\Delta m^2_{13}|=3\times 10^{-3}$~eV$^2$, 
$\sin^2\theta_{\rm atm}=1/2$
are precisely known. Fig.~\ref{measure_low}(top,right)
depicts $\chi^2-\chi^2_{\rm MIN}$ as a function of $|U_{e3}|^2$ 
corresponding to 120~kton-years\footnote{This corresponds to
six years of running with the current NuMI beam configuration and
a 20 kton detector. With a proton driver, however, the same amount of
data can be collected in 1.5 years. This will become crucial later.} 
of ``data'' collected with 
a neutrino beam (as defined earlier, the neutrino (antineutrino) 
beam consists predominantly of 
$\nu_{\mu}$ ($\bar{\nu}_{\mu}$)). Note that, while the data were simulated with
$\Delta m^2_{13}=+3\times 10^{-3}$~eV$^2$ and $|U_{e3}|^2=0.008$, a different
solution, with the same goodness of fit, is found for $\Delta m^2_{13}
=-3\times 10^{-3}$~eV$^2$, $|U_{e3}|^2=0.015$.\footnote{It is important
to reemphasize that $|\Delta m^2_{13}|$
is {\sl not} a fit parameter. It is assumed to be known from different
sources, possibly the study of the $\nu_{\mu}\rightarrow\nu_{\mu}$
disappearance channel in the off-axis experiment!} This implies that
if the neutrino mass hierarchy is not known, instead of obtaining a precise
measurement of $|U_{e3}|^2=0.008\pm 0.0025$ (these are two sigma error bars), 
one is forced to quote a less precise (very non-Gaussian) 
measurement: $0.0055<|U_{e3}|^2<0.018$ at the 
two-sigma confidence level.   

The situation can be improved significantly if, after running with
the neutrino beam, one runs with the antineutrino beam. 
Fig.~\ref{measure_low}(top,right) depicts $\chi^2-\chi^2_{\rm MIN}$ 
as a function of $|U_{e3}|^2$ corresponding to 300 kton-years of 
``data'' collected with the antineutrino beam. As mentioned before, 
one is required to run much longer with the antineutrino beam in 
order to obtain a statistical significance comparable to the one
obtained with the neutrino beam. One should readily note that 300 kton-years
would correspond to 15 years (!) of running with the current NuMI beam configuration
and a 20 kton off-axis detector. Here the presence of a proton driver becomes 
vital: it increases the beam intensity by a significant factor (nominally four),
and reduces the 15 years to a bearable 3.74 years, such that the combined neutrino
plus antineutrino beam time is slightly longer than 5 years.
 
Again, the same behavior as
before is observed: one obtains two distinct values of $|U_{e3}|^2$ depending
on the assumption regarding the neutrino mass hierarchy, with a significant
difference: this time the effect is ``reversed.'' The reason for this is
simple: with the neutrino beam, the inverted hierarchy reduces the
$\nu_e$ appearance signal compared to the normal hierarchy and, therefore, 
in order to correctly fit the data, a larger value of $|U_{e3}|^2$ 
(compared to the one obtained with the normal hierarchy) is preferred. In the
case of the antineutrino beam, the inverted hierarchy enhances the
$\bar{\nu}_e$ appearance signal, and a smaller value of $|U_{e3}|^2$ is
preferred. This allows one to separate the two signs of $\Delta m^2_{13}$
if the information obtained with both beams is combined. This is what is
done in Fig.~\ref{measure_low}(bottom,left). Note that in this case
the ``wrong'' model is about sixteen units of $\chi^2$ away from the ``right''
model. It is also curious to note that, even with the wrong hypothesis,
a similar measurement of $|U_{e3}|^2$ is obtained. This coincidence, which
will not be considered too relevant, is a consequence of the fact that
the data with the neutrino and antineutrino beams ``pull'' the
measured $|U_{e3}|^2$ in opposite direction, and their combination meets 
somewhere ``in the middle.''

Finally, in order to determine how well the two different signs of
$\Delta m^2_{13}$ can be separated, Fig.~\ref{measure_low}(bottom,right)
depicts $\chi^2_{\rm PLUS}-\chi^2_{\rm MINUS}$ as a function of the
input value of $|U_{e3}|^2$, plus the input $\Delta m^2_{13}>0$. Note
that for $|U_{e3}|^2\gtrsim 2\times
10^{-3}$, a $\chi^2$ separation of more than 
two units can be obtained. One can turn this into an exclusion of the
``wrong'' sign by noting that, for an average experiment, $\chi^2=2$
for the ``correct'' hierarchy, 
if one combines the data obtained with the two beams (2 is the number
of degrees of freedom in this case). This implies that, for 
$|U_{e3}|^2=0.01$, the wrong hypothesis yields $\chi^2\simeq 24$, which is 
excluded at more than four sigma. A three sigma confidence level
determination of the neutrino mass hierarchy
would be obtained at $|U_{e3}|^2\simeq 0.005$.  

In summary, if the LMA solution is excluded by KamLAND \cite{Kam}
(or, perhaps, a
solar neutrino experiment, like Borexino \cite{Borexino}), future long baseline
neutrino experiments can only hope to measure the magnitude of the 
$U_{e3}$ element of the neutrino mixing matrix, and determine the
neutrino mass hierarchy. On the other hand, both the estimated sensitivity reach
and the measurement of $|U_{e3}|^2$ are very ``clean,'' in the sense that
they are not clouded by other physical effects (this will become clear
after the next couple of sections). Furthermore, it is important to emphasize
that both of these tasks are of the utmost importance, and are already enough to
justify pursuing this kind of activity.  

\subsection{LMA Solution with Spectral Distortions at KamLAND}

If the best fit point to the current solar data \cite{solar_fits} 
is close to reality, the
KamLAND reactor neutrino experiment will be able to not only observe 
a depletion of the reactor antineutrino flux, but also determine the values
of $\Delta m^2_{12}$ and $\sin^2\theta_{\odot}$ with very good precision 
\cite{Kam,Kam_sim,Kam_sim2,Kam_sim3}. 

\begin{figure}
\centerline{\epsfxsize 13cm \epsffile{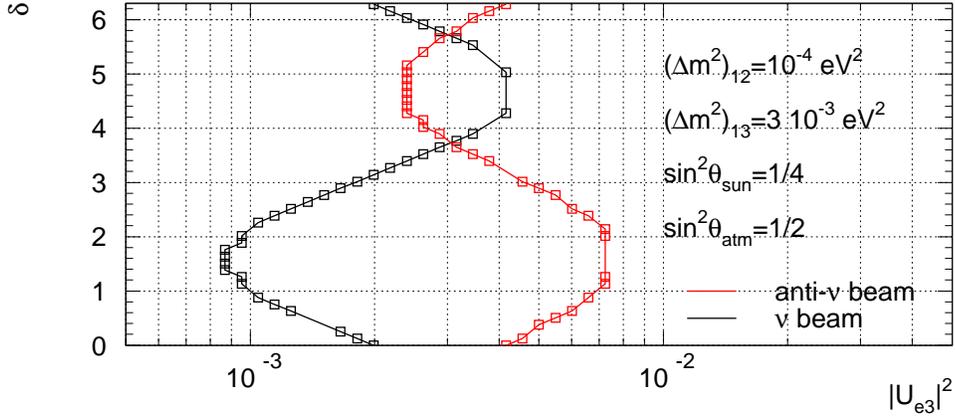}}
\caption{Three sigma sensitivity for observing a $\nu_{\mu}\rightarrow\nu_e$ signal
in the $(|U_{e3}|^2\times\delta)$-plane, 
after 120 kton-years of neutrino-beam running (black, darker line)
or 300 kton-years of antineutrino-beam running (red, lighter line). 
$\Delta m^2_{13}=+3\times10^{-3}$~eV$^2$,
$\sin^2\theta_{\rm atm}=1/2$, $\Delta m^2_{12}=1\times 10^{-4}$~eV$^2$, 
$\sin^2\theta_{\odot}=1/4$. }
\label{sens_lma}
\end{figure}

This being the case, in this section we address the issue of determining
the neutrino mixing parameters $|U_{e3}|^2$ and $\delta$ at the off-axis 
experiment. As in the previous section, we start by determining the 
sensitivity of the off-axis experiment and 120~kton-years running of the 
neutrino beam to observing an excess of $\nu_e$-like events. 
Fig.~\ref{sens_lma} depicts the three sigma sensitivity in the ($\delta\times
|U_{e3}|^2$)-plane for $\Delta m^2_{13}=3\times 10^{-3}$~eV$^2$\footnote{As 
before, the sensitivity is close to optimal for $\Delta m^2_{13}=(2-4)\times 
10^{-3}$~eV$^2$. One should keep in mind that if the atmospheric mass-squared
difference turns out to be significantly different from this range, a different
beam configuration has to be consider in order to optimize the appearance signal
in the off-axis detector.} for
120 (300)~kton-years of running with the (anti)neutrino beam. 
The sensitivity depends significantly on the CP-odd phase $\delta$, and,
as expected, the sensitivity is best for $\delta\sim\pi/2$ in the
case of running with a neutrino beam ($\delta\sim3\pi/2$ for the 
antineutrino beam), where the
``interference'' between the ``CP-odd term'' and the ``$|U_{e3}|^2$ term''
is constructive ({\it i.e.,}\/ one observes more events) and worse at
$\delta\sim3\pi/2$, where the ``interference'' is destructive. For smaller 
values of $\Delta m^2_{12}$, the `z-shape' and `s-shape' observed in 
Fig.~\ref{sens_lma} degenerate into vertical straight lines, such that the 
sensitivity will no longer depend on the CP-odd phase. 

\begin{figure}
\centerline{\epsfxsize 14.2cm \epsffile{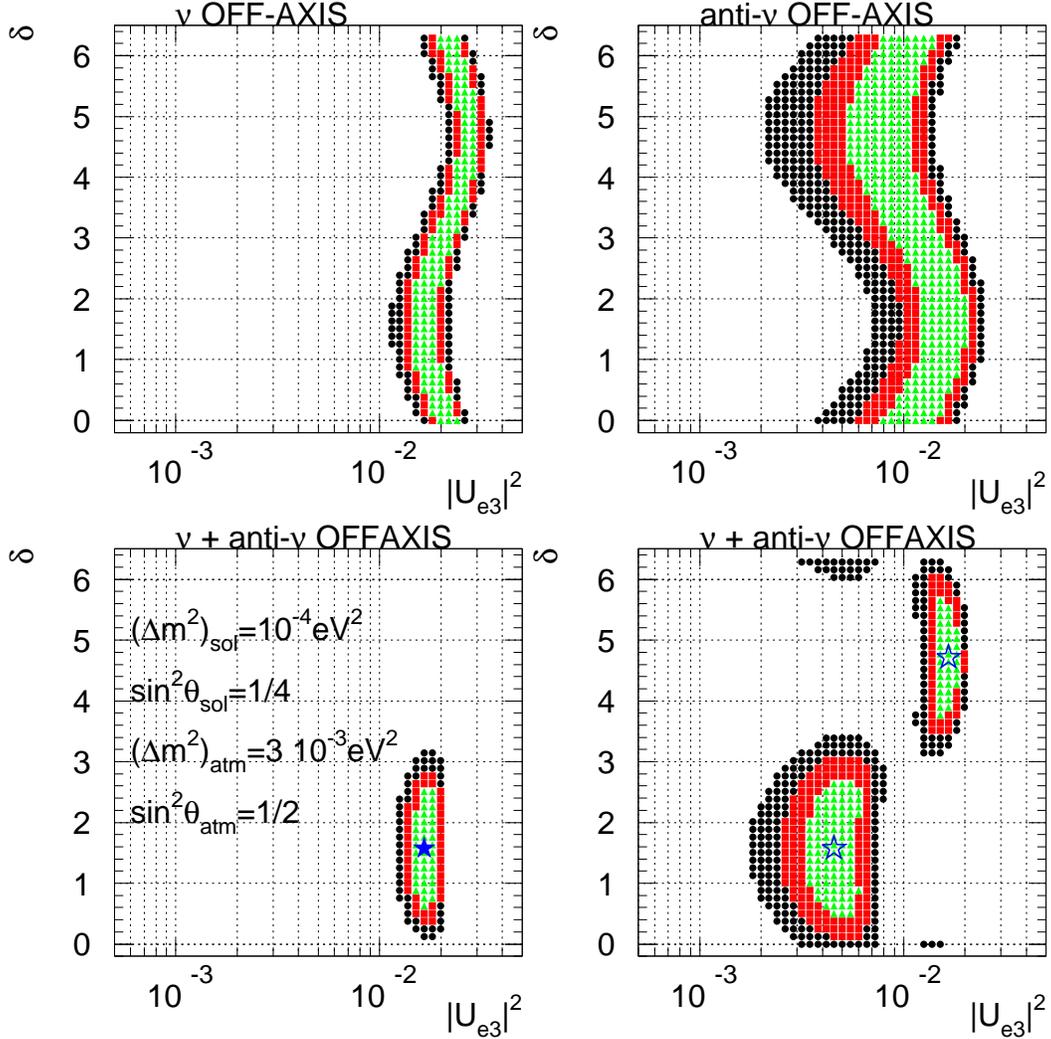}}
\caption{top -- One, two, and three sigma measurement contours in the 
$(|U_{e3}|^2\times\delta)$-plane, after 120 kton-years of neutrino-beam running 
(left) or 300 kton-years of antineutrino-beam running (right). The simulated data 
is consistent with $|U_{e3}|^2=0.017$ and $\delta=\pi/2$. bottom,left -- same as above,
after the two data sets are combined. The solid star indicates the simulated input. 
bottom,right -- same as before, for different simulated data points (indicated by the stars).
$\Delta m^2_{13}=+3\times10^{-3}$~eV$^2$,
$\sin^2\theta_{\rm atm}=1/2$, $\Delta m^2_{12}=1\times 10^{-4}$~eV$^2$, 
$\sin^2\theta_{\odot}=1/4$.}
\label{measure_lma}
\end{figure}

If a signal is observed, one can attempt to determine the
mixing parameters $|U_{e3}|^2$ and $\delta$.
Similar to what was done in the previous section, the atmospheric
parameters $\Delta m^2_{13}=3\times10^{-3}$~eV$^2$,
$\sin^2\theta_{\rm atm}=1/2$ will be assumed known
with infinite precision, and the same will now hold for the solar parameters
$\Delta m^2_{12}=1\times 10^{-4}$~eV$^2$, $\sin^2\theta_{\odot}=1/4$. 
Furthermore, we will also assume that
the neutrino mass hierarchy is known.\footnote{It may turn out,
for example, that table top experiments \cite{double_beta}
or the observation of supernova neutrinos \cite{supernovae}
will be able to measure the neutrino mass hierarchy.} This is done in order to 
not cloud the results presented here. Fig.~\ref{measure_lma}(top,left) 
depicts the one, two, and three sigma measurement contours in the ($|U_{e3}|^2
\times\delta$)-plane obtained after 120~kton-years running with the neutrino 
beam.
The simulated data are consistent with $|U_{e3}|^2=0.017$ and $\delta=\pi/2$. 
One can readily note that while
$|U_{e3}|^2$ can be measured with reasonable precision, virtually nothing can
be said about $\delta$. Furthermore, the fact that $\delta$ is not known implies
that a measurement of $|U_{e3}|^2$ irrespective of $\delta$ 
is in fact less precise than what can be obtained
if the solar parameters are not in the LMA region. 

In order to improve on this picture, it is imperative to prolong our 
``experiment'' and take data with the antineutrino beam as well. 
Fig.~\ref{measure_lma}(top,right) depicts the one,
two, and three sigma measurement contours in the ($|U_{e3}|^2\times\delta$)-plane
obtained after 300~kton-years running with the antineutrino beam (as mentioned
before, the longer running time is required in order to compensate for the 
``less efficient'' antineutrino beam). Again, $|U_{e3}|^2$ can be measured 
with some precision and nothing can be said about $\delta$. A comparison
of the two figures hints that a combined analyses may prove more fruitful. 
This is the case because while
the neutrino beam yields a ``z-shaped'' measurement contour, the antineutrino
beam yields an ``s-shaped'' contour. The reason for this is that the number of
$\nu_{\mu}\rightarrow\nu_e$ induced events 
is larger for $\delta=\pi/2$ and smaller for $\delta=3\pi/2$.
Therefore, the measurement will choose larger values of $|U_{e3}|^2$ at around
$\delta=3\pi/2$ in order to compensate for this small suppression. On the other
hand, the number of $\bar{\nu}_{\mu}\rightarrow\bar{\nu}_e$ induced 
events is smaller at
$\delta=\pi/2$ and larger at $\delta=3\pi/2$, and the opposite phenomenon
is observed.

Fig.~\ref{measure_lma}(bottom,left) depicts the result of measuring $|U_{e3}|^2$
and $\delta$ using the combined neutrino and antineutrino beam ``data.'' The 
situation is significantly improved, and now, a three sigma measurement of 
$\delta\neq 0$ can be performed. Fig.~\ref{measure_lma}(bottom,right) is
similar to Fig.~\ref{measure_lma}(bottom,left), except that different input
values of $|U_{e3}|^2,\delta$ are chosen. As expected, the quality of the
measurement is marginally worse for $\delta=3\pi/2$ (where $\delta$ 
is consistent
with zero at the three sigma level), and deteriorates as $|U_{e3}|^2$ decreases.
For $\delta=\pi/2$, one cannot determine that $\delta\neq 0$ or $\pi$ 
({\it i.e.,}\/ no CP-violation) at the
two sigma level if $|U_{e3}|^2\lesssim 0.004$. It should always be kept in mind that
the situation deteriorates for smaller values of $\Delta m^2_{12}$.   

It is worthwhile to note that
an alternative (perhaps more standard) way to look at CP violation would be to 
construct an ``asymmetry-like'' parameter ({\it e.g.},\/ something
proportional to $P(\nu_{\mu}\rightarrow \nu_e)-P(\bar{\nu_{\mu}}
\rightarrow\bar{\nu}_e)$). This approach will not yield a better measurement
of the oscillation parameters $\delta$ and $|U_{e3}|^2$ (as we are using all
the experimental data and perform a ``global fit''). Furthermore, it is not
even an appropriate ``direct detection'' of CP-violation, given that our
neutrino and antineutrino beams are not pure (especially the antineutrino 
beam) {\sl and} matter effects contribute a significant amount of ``fake'' 
CP-violation. 

In summary, if KamLAND observes a suppressed and distorted reactor antineutrino
spectrum, long baseline neutrino experiments can potentially measure both
the magnitude of the $U_{e3}$ element of the neutrino mixing matrix 
and the CP-odd Dirac phase.
Note that attempting to measure the CP-odd phase is not ``optional'' -- the
fact that it is unknown introduces a substantial uncertainty on determining 
$|U_{e3}|^2$. In order to disentangle the two unknown parameters, it is crucial 
to take data using both a neutrino and an antineutrino beam.\footnote{Another 
possibility would be to use two different detectors at different positions 
and/or a different beam with a different energy spectrum. We do not consider 
this possibility in this study.} 
%Finally, it is important to comment 
%that we would profit from a `cleaner'' antineutrino beam -- note that
%the `z-shape' obtained with the neutrino beam is more pronounced than the
%`s-shape' obtained with the antineutrino beam (see Fig.~\ref{measure_lma}).

\subsection{HLMA Solution, no Spectral Distortions at KamLAND}

If $\Delta m^2_{12}\gtrsim 2\times 10^{-4}$~eV$^2$ (note that this possibility
is not currently excluded by solar neutrino data \cite{solar_fits}), 
KamLAND will not be sensitive to the very rapid oscillatory pattern, and 
will only be able to observe an overall suppression of the solar neutrino 
flux \cite{Kam_sim,Kam_sim3}. In this case, the 
mixing angle $\sin^2\theta_{\odot}$ can be measured with some precision
by determining the overall suppression factor, but the value of 
$\Delta m^2_{12}$ will only be constrained to be larger than some lower limit.
An upper limit will be provided by either future solar data or by the current
CHOOZ data \cite{CHOOZ}. In order to be conservative, we will consider 
the latter case, and assume that $\Delta m^2_{\odot}\lesssim
7\times 10^{-4}$~eV$^2$ for large solar angles.

If this scenario turns out to be correct, precise
measurements of the atmospheric parameters and the solar mixing angle 
will probably be available, while 
$|U_{e3}|^2$, $\delta$ and the precise value of 
$\Delta m^2_{12}$ will remain unknown.\footnote{A short-KamLAND or long-CHOOZ 
reactor experiment would certainly resolve this issue \cite{miniKam}. Such
an experiment has not been proposed yet (see, however, \cite{miniKam_exp}).} 

What are the consequences of having a very large but poorly measured 
$\Delta m^2_{12}$? The biggest consequence, perhaps, is that even for very
small values of $|U_{e3}|^2$, a significant amount of $\nu_e$-like events
will be observed. This implies that the ``sensitivity to $|U_{e3}|^2$,''
as discussed in the two previous sections is not a particularly meaningful
quantity to study. Furthermore,
as one may fear, this will also lead to a $\Delta m^2_{12}$ versus
$|U_{e3}|^2$ ``confusion,'' (this was already alluded to in \cite{Kam_sim3}) 
similar to the one observed between $|U_{e3}|^2$
and $\delta$ in the previous section (and which continues to exist here, of 
course). In other words, a moderate $\Delta m^2_{12}$ and a large 
$|U_{e3}|^2$ will yield as many events as a large 
$\Delta m^2_{12}$ and a small $|U_{e3}|^2$. We already learned from the
previous two sections that one will be required to run both the neutrino and
the antineutrino beams (and accumulate enough statistics with both) in 
order to try to disentangle the three parameters.
One should keep in mind that,
while the situation is rather confusing, the number
of observed events is going to be large for very large values of the solar
mass-squared difference.

In order to address what kind of measurement one may be able to perform under
these conditions, we simulate data for $\delta=\pi/2$, $|U_{e3}|^2=0.012$,
$\Delta m^2_{12}=4\times 10^{-4}$~eV$^2$, 
$\Delta m^2_{13}=3\times 10^{-3}$~eV, 
 $\sin^2\theta_{\rm atm}=1/2$, and $\sin^2\theta_{\odot}=1/4$, 
assuming 120~kton-years of neutrino beam running and 300~kton-years of 
antineutrino beam running. As before, we assume during the data analysis that 
the atmospheric parameters, the neutrino mass hierarchy and the solar 
angle are known with infinite precision.

\begin{figure}
\centerline{\epsfxsize 13cm \epsffile{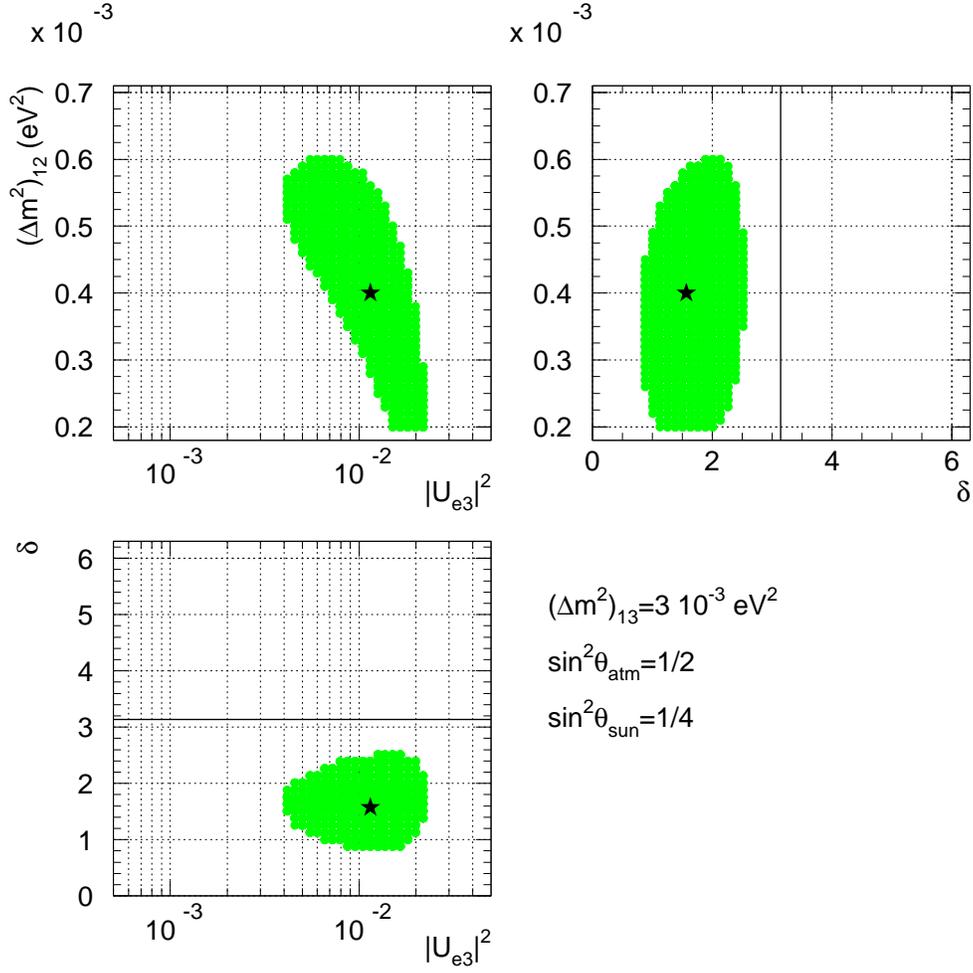}}
\caption{ Projections of the three sigma measurement surface in the 
$(|U_{e3}|^2\times\delta\times\Delta m^2_{12})$-space, after 120 kton-years of neutrino-beam 
running and 300 kton-years of antineutrino-beam running. The simulated data 
is consistent with $|U_{e3}|^2=0.012$, $\delta=\pi/2$ and $\Delta 
m^2_{12}=4\times 10^{-4}$~eV$^2$. 
$\Delta m^2_{13}=+3\times10^{-3}$~eV$^2$,
$\sin^2\theta_{\rm atm}=1/2$, 
$\sin^2\theta_{\odot}=1/4$.}
\label{measure_hlma}
\end{figure}

The results of the three parameter fit are presented in Fig.~\ref{measure_hlma},
where we plot the three two-dimensional projections of the three sigma 
surface in the 
($\Delta m^2_{12}\times|U_{e3}|^2\times\delta$)-space. Many comments are in
order. First of all, one should note that the solar mass difference cannot be
measured with any reasonable precision -- it is only slightly better known
than before, namely, it lies somewhere between $2\times 10^{-4}$~eV$^2$
(the KamLAND bound) and $6\times 10^{-4}$~eV$^2$ (slightly better than 
the CHOOZ bound). The top 
left-hand panel depicts the $\Delta m^2$ versus $|U_{e3}|^2$ confusion alluded
to earlier quite well. It is curious to note however, that the capability to 
decide whether $|U_{e3}|^2$ is nonzero or not is not weak -- 
the loose constraints on $\Delta m^2_{12}$ are already 
enough to guarantee that $|U_{e3}|^2\gtrsim 4\times 10^{-3}$. Most importantly,
perhaps, at the three sigma level there is solid
evidence that $\delta\neq 0,\pi$. The reason for this is that,
for $\Delta m^2_{12}=4\times 10^{-4}$ and $\delta=\pi/2$, 
$\bar{\nu}_{\mu}\rightarrow \bar{\nu}_e$ transitions are very suppressed, 
and a zero CP-odd phase would yield far too many $\bar{\nu}_e$-like events 
when the antineutrino beam is on.  

In summary, for very large values of $\Delta m^2_{12}$, KamLAND 
will not be able to measure the solar mass-squared difference 
with reasonable precision. Under these circumstances, long-baseline
neutrino experiments are required to measure $\Delta m^2_{12}$ along
with $|U_{e3}|^2$ and $\delta$. We find that, if $|U_{e3}|^2$ is large
enough, there is a reasonable chance that CP violation can be observed,
while $|U_{e3}|^2$ can be measured rather poorly. The 
measurement of $\Delta m^2_{12}$ is even less precise. As before, 
accumulating enough statistics
with both the neutrino and antineutrino beams is required,
and, particularly on this case, one 
would profit immensely from other measurements with different distances 
and/or baselines (see, for example, \cite{JHF}).   

\section{Summary and Conclusions}

The ultimate goal for the next generation of neutrino experiments
will be to tie some of the ``loose ends'' of neutrino masses and mixings \ie, \ 
determine the neutrino mass hierarchy, measure
(or further constraint) the ``connecting mixing angle''
($|U_{e3}|^2$), and explore leptonic CP violation (if the solution to
the solar neutrino puzzle lies in the LMA region). 

Candidate next generation experiments include shooting an intense conventional
muon-type neutrino beam towards a detector which is located conveniently away from
the main beam direction. 
``Off-axis beams'' have several advantages with respect
to their on-axis counterparts. They are more intense,
narrower and lower energy beams, 
``void'' of a large high energy tail, and  provide clean access to 
$\nu_e$-appearance, given the existence of a suitable detector with good electron
identification capabilities.
In order to fully take advantage of such tools, however, we
have argued that two different beams (a predominantly $\nu_{\mu}$ and a
predominantly $\bar{\nu}_{\mu}$ beam) are essential. ``Antineutrino beams,'' however,
provide an extra experimental challenge: the antineutrino
cross sections are suppressed with respect to the neutrino ones, such that,
in order to obtain a statistically comparable antineutrino and neutrino data set,
a sizeable amount of running time is required. The solution we find to this
issue is a more intense proton source, along with a large (but realistically sized) detector.

Specifically, 
we studied how well experiments in the NuMI beamline with a proton driver 
upgrade plus an off-axis detector can address these issues.
In order to properly estimate the experimental response ({\it e.g.,}\/ 
signal efficiency as well
as beam and detector-induced backgrounds), we performed a realistic simulation of the NuMI
beam plus the response of a 
highly segmented iron detector, followed by a detailed ``data'' analysis.
The combination of a new proton driver and the off-axis detector
using the NuMI beamline can, in five years, improve considerably 
our knowledge of the neutrino sector.

To properly assess the capabilities of such a set up, it is crucial to explore all 
the different physics scenarios, which will be (hopefully) distinguished by the
current KamLAND reactor experiment. For different values of the solar mass-squared
difference, we obtain different results for a five year program with an upgraded
NuMI beam and a 900~km long baseline off-axis experiment:

\begin{enumerate}
\item KamLAND does not observe a suppression of the reactor neutrino flux --
$|U_{e3}|^2$ can be measured with very good precision and the neutrino mass pattern
can be established, as long as $|U_{e3}|^2\gtrsim \rm few \times 10^{-3}$.
We emphasize that even in this ``less fortunate scenario''
key information regarding neutrino physics will be obtained.

\item KamLAND sees a distortion of the reactor neutrino
spectrum -- one should be capable of measuring $|U_{e3}|^2$ with good precision
and obtaining a rather strong hint for CP violation,
as long as $|U_{e3}|^2\gtrsim \rm few \times 10^{-3}$, 
$\delta$ close to either $\pi/2$ or $3\pi/2$. 
As a simplifying assumption, we assumed that the neutrino mass hierarchy would be
already determined by other means.

\item KamLAND sees an oscillation signal but is not able to measure $\Delta m^2_{\odot}$ --
one should be capable of measuring $|U_{e3}|^2$ with some precision
and obtaining a strong hint for CP violation
as long as $|U_{e3}|^2\gtrsim 10^{-2}$, $\delta$
close to either $\pi/2$ or $3\pi/2$, even if $\Delta m^2_{\odot}$ is poorly known.
$\Delta m^2_{\odot}$ cannot be measured with any reasonable precision. Again,
as a simplifying assumption, we assumed that the neutrino mass hierarchy would be
already determined by other means.
\end{enumerate}

It is also important to stress that, as discussed here in some detail, 
order 2~GeV neutrinos and a 900~km baseline are appropriate to measure fake as
well as genuine CP violation  ({\it i.e.,}\/ both the neutrino mass hierarchy and
the Dirac phase $\delta$ of the neutrino mixing matrix). Furthermore, we have argued that
similar results should be obtained for different baselines (as long as the off-axis
distance is appropriately modified) and for different values of the atmospheric mass-squared
difference (see Fig.~\ref{fom}). In particular, 
the latter implies that one need not wait until a very precise measurement 
of $\Delta m^2_{\rm atm}$ is made in order to decide where the off-axis detector is to
be located.

We conclude by comparing some of the results obtained here with similar
studies performed for a future JHF to Kamioka neutrino beam \cite{JHF}.
A neutrino beam from the  JHF-proton source~\cite{JHF}  aimed at the
Super-Kamiokande detector~\cite{SK} is a neutrino project with a 295~km baseline aimed
to start in 2007-2008.
Similar to the NuMI-off-axis project, the physics goals are to measure with
an order of magnitude better precision (compared to MINOS and CNGS estimates \cite{bar})
the atmospheric parameters ($\delta(\Delta m_{\rm atm}^2)\lesssim 10^{-4}$~eV$^2$ and
$\delta(\sin^2 2\theta_{\rm atm})\lesssim 0.01$),
confirm $\nu_\mu \leftrightarrow \nu_\tau$-oscillations or discover 
sterile neutrinos by measuring the neutral current event rate, and to improve by
a factor of 20 the sensitivity to $\nu_\mu \to \nu_e$-appearance.
After a five year program, the JHF-Kamioka program
should be able to exclude, at the 90\% confidence level,
$\nu_\mu \to \nu_e$ transitions for $|U_{e3}|^2>0.0015$, while at NuMI
with a 20~Kton  off-axis detector we will exclude, at the
two sigma confidence level, $|U_{e3}|^2>0.00085~(0.0015)$
with~(without) an upgrade proton driver (assuming $\Delta m^2_{\odot}\ll 10^{-4}$~eV$^2$
and a normal neutrino mass hierarchy). The main difference between the two programs
should come from the longer baseline proposed here, which allows the NuMI 
off-axis experiment (but not the JHF-Kamioka program) to cleanly 
try to address the neutrino mass hierarchy.

\subsection*{Acknowledgements}
\noindent
We thank Maury Goodman and Doug Michael for comments and suggestions, and Robert Hatcher
and Gokhan Unel for providing technical support regarding the various simulation
tools. We also thank the participants of the ``Proton Driver Physics Study Group'' based at
Fermilab for useful discussions. 
GB and AdG were supported by the U.S.~Department of Energy
Grant DE-AC02-76CHO3000 while MS and MV
were supported in part by grants from the Illinois Board of
Higher Education, the Illinois Department of Commerce and Community
Affairs, the National Science Foundation, and the U.S. Department of
Energy.

\end{document}